\documentclass[manuscript, screen]{acmart}
\AtBeginDocument{%
  }

\usepackage[ruled,vlined,linesnumbered]{algorithm2e} 

\usepackage{multirow}
\usepackage{pifont}
\usepackage{graphicx}
\usepackage{amsmath}
\usepackage{comment}
\usepackage{verbatim}
\usepackage{xcolor}
\usepackage{colortbl}

\usepackage{acronym}
\usepackage{fancyhdr}
\usepackage[normalem]{ulem}
\useunder{\uline}{\ul}{}
\usepackage{balance}
\usepackage{booktabs}
\usepackage{arydshln} 

\begin{document}


\title{Optimizing VLP-aligned Multimodal Intent Representation with Correct Visual Instantiation for Zero-Shot Composed Image Retrieval}

\author{Xuri Ge}

\affiliation{%
  \institution{School of Artificial Intelligence, Shandong University}
  \city{Jinan}
  \state{Shandong}
  \country{China}
}
\email{xur.ge@sdu.edu.cn}

\author{Chunhao Wang}
\affiliation{%
  \institution{School of Computer Science and Technology, Shandong University}
  \city{Qingdao}
  \state{Shandong}
  \country{China}
}
\email{202415164@mail.sdu.edu.cn}

\author{Junchen Fu}
\affiliation{%
  \institution{School of Computing Science, University of Glasgow}
  \city{Glasgow}
  \country{UK}
}
\email{2628551f@student.gla.ac.uk}

\author{Haokun Wen}
\affiliation{%
 \institution{School of Computer Science and Technology, Harbin Institute of Technology (Shenzhen)}
 \city{Shenzhen}
 \country{China}}
\email{haokunwen@outlook.com}

\author{Zhiwei Xu}
\affiliation{%
  \institution{School of Artificial Intelligence, Shandong University}
  \city{Jinan}
  \state{Shandong}
  \country{China}
}
\email{zhiwei_xu@sdu.edu.cn}

\author{Ying Zhou}
\affiliation{%
  \institution{School of Artificial Intelligence, Shandong University}
  \city{Jinan}
  \state{Shandong}
  \country{China}
}
\email{yingzhou@sdu.edu.cn}

\author{Zhumin Chen }
\affiliation{%
  \institution{School of Artificial Intelligence, Shandong University}
  \city{Jinan}
  \state{Shandong}
  \country{China}
}
\email{chenzhumin@sdu.edu.cn}

\author{Pengjie Ren }
\affiliation{%
  \institution{School of Computer Science and Technology, Shandong University}
  \city{Qingdao}
  \state{Shandong}
  \country{China}
}
\email{renpengjie@sdu.edu.cn}

\author{Zhaochun Ren}
\affiliation{%
  \institution{Leiden University}
  \city{Leiden}
  \country{The Netherlands}}
\email{z.ren@liacs.leidenuniv.nl}

\author{Xin Xin}
\authornote{corresponding author.}
\affiliation{%
  \institution{School of Artificial Intelligence, Shandong University}
  \city{Jinan}
  \state{Shandong}
  \country{China}
}
\email{xinxin@sdu.edu.cn}
\renewcommand{\shortauthors}{Xuri Ge et al.}

\begin{abstract}
Zero-shot composed image retrieval (ZS-CIR) aims to retrieve a target image from a reference image and a modification text without paired supervision, typically by encoding composed queries as text-dominant representations within the image-text matching space of vision-language pre-trained models (VLPs). However, queries reconstructed by mainstream visual pseudo-word learning or multimodal large language model (MLLM) based target reasoning paradigms often remain incompatible with the native representation space of VLPs.  This incompatibility arises because visual pseudo-words may introduce reference noise with coarse text fusion, while MLLM-generated descriptions are verbose and weakly visually grounded, both deviating from the visually grounded, retrieval-oriented text space learned by VLPs.In this paper, we propose a unified ZS-CIR framework (named VMIR-CVI) to reconstruct multimodal composite queries from two complementary perspectives for optimizing VLP-compatible multimodal intent representation.  First, it reasons and converts the multimodal intent into a unified textual description, aligning with the native text space of the VLP backbones to produce more retrieval-compatible textual queries. Second, it reconstructs the query representation with correctly decoupled visual instance cues, reducing reference noise while preserving target-relevant content. Specifically, a VLP-aligned Multimodal Intent Reasoning (VMIR) module injects few-shot VLP-style exemplars into chain-of-thought prompts, guiding the MLLM to generate target-consistent intent queries. A Training-free Visual Instance Disentanglement (TVID) module decouples fine-grained visual instances from global reference features without additional optimization. Finally, a lightweight Hybrid-modal Intent Alignment and Fusion (HIAF) module integrates the reasoned textual intent and disentangled visual cues into a unified hybrid-modal representation for robust ZS-CIR.
Extensive experiments on three CIR benchmarks, namely CIRR, CIRCO and FashionIQ,  show that VMIR-CVI significantly outperforms existing baselines and achieves new state-of-the-art performance. Code and trained models will be publicly released.

\end{abstract}



\maketitle
\begin{figure}[t] 
    	\centering
    	\includegraphics[width=0.7\linewidth]{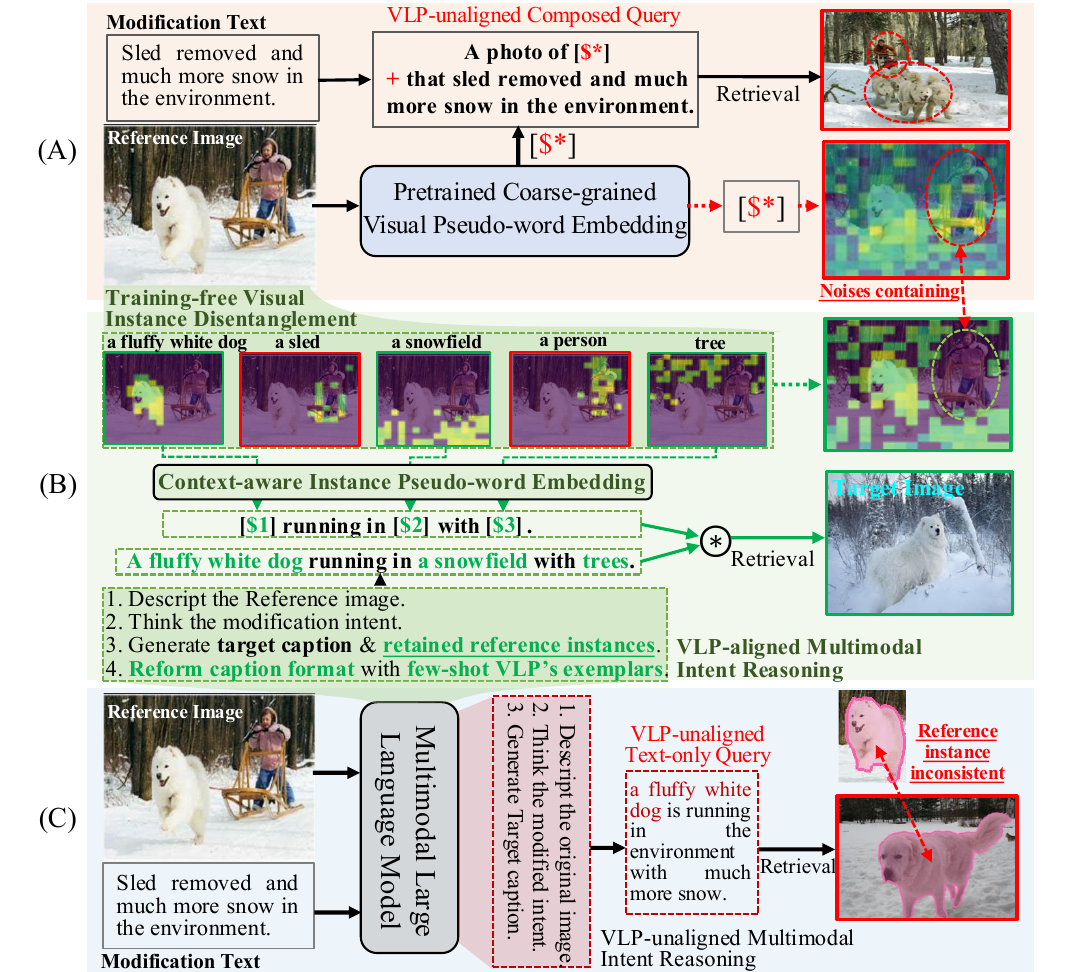}
        \caption{Compared with the mainstream zero-shot CIR models based on the visual pseudo-word embedding model (A) and based on MLLM's multimodal intent reasoning model (C), we introduce a VLP-aligned multimodal intent representation with correct visual instantiation (VMIR-CVI) model (B).
        In this way, we can effectively integrate fine-grained disentangled correct reference instances into VLP-aligned query, which can significantly avoid the interference of reference visual noise in mechanism (A) and the inconsistency of visual reference instances in mechanism (C), as well as unfriendly retrieval paradigm aligned with VLP backbones.
        }  
    	\label{fig:motivation}
\end{figure} 
\section{Introduction}

Traditional image retrieval methods based on visual similarity~\cite{rehman2012content} and text--image semantic matching~\cite{cao2022image} have achieved strong performance in simple and well-defined retrieval scenarios. However, many real-world search applications require users to specify a target not by an image alone or a text query alone, but by modifying a reference image with natural language descriptions. 
This has directly driven the development of the Composed Image Retrieval (CIR) task~\cite{vo2019composing,xing2025context,byun2025ma}, whose goal is to retrieve a target image that satisfies the composed intent jointly expressed by a reference image and a modification text. 
Compared with conventional image retrieval, CIR is fundamentally more challenging because the model must not only understand the visual content of the reference image, but also infer how the textual modification changes, i.e., which visual attributes or instances should be preserved, removed, or transformed. Owing to this unique ability to support flexible and fine-grained user intent expression, CIR has become increasingly important in applications such as e-commerce~\cite{wu2024personalized} and interactive search~\cite{xie2020modeling,luo2025imagescope}.

While CIR enables more flexible user interactions, most traditional CIR methods~\cite{wen2024simple,psomas2025instance,li2026habit} are formulated in a fully supervised manner and rely on large-scale annotated triplets of the form $(\text{reference image}, \text{modified text}, \text{target image})$. These methods learn multimodal intent composition by directly optimizing the match between a composed query and its corresponding target image, often achieving strong performance under closed training distributions. However, collecting such triplet annotations is expensive and labor-intensive, and the resulting models often generalize poorly to unseen compositions or domains without further supervision. 
To alleviate this dependency, recent research has increasingly focused on \textit{zero-shot CIR}, where models must generalize to unseen compositions without paired supervision, greatly enhancing practicality and scalability in open and dynamic scenarios.

Existing zero-shot CIR approaches aim to perform composed image retrieval without annotated supervision, which are mainly built upon existing \textit{vision–language pretrained models (VLPs)} \cite{radford2021learning,li2022blip}. Because large-scale pre-training provides strong cross-modal alignment and generalization ability. 
Under a frozen VLP backbone, the core challenge of zero-shot CIR is to reconstruct a unified and retrievable representation for multimodal composed queries. Such a representation should satisfy two requirements simultaneously: on the one hand, it should be compatible with the retrieval space of the given VLP and follow the contextual expression style of its native text space; on the other hand, it should effectively integrate multimodal composed intent, especially the correct visual reference and the textual modification, to capture user intent more accurately and comprehensively.
Therefore, a central challenge for improving zero-shot CIR is \textit{how to better optimize VLP-aligned multimodal intent representations of composed queries for VLP backbones} to match target image.

Early representative studies (shown in Figure~\ref{fig:motivation}(A))~\cite{saito2023pic2word,gu2024language,li2025rethinking} follow the line of \emph{visual pseudo-word learning}, where the reference image is projected into one or multiple pseudo-text tokens and concatenated with the modification text to form a unified query under the frozen VLP backbone. This paradigm directly reuses the pretrained text encoder for composed retrieval, effectively converting the multimodal query into a text-compatible representation. For example, Pic2Word~\cite{saito2023pic2word} employs a lightweight adapter to map the global image feature into a pseudo-word, enabling unsupervised visual--textual fusion for zero-shot retrieval. Subsequent methods~\cite{wang2025mapping,lin2024fine} further improve this line by learning more expressive visual pseudo-word representations. However, they still reconstruct the composed query representation through coarse concatenation of embedded reference visual content and modification text within the VLP space.

With the rapid development of multimodal large language models (MLLMs), recent studies (Figure~\ref{fig:motivation}(C))~\cite{sun2023training,karthik2023vision,yang2025explainable,yang2024ldre,tang2025reason} instead leverage their strong reasoning and instruction-following capabilities to explicitly infer composed retrieval intent. Rather than learning pseudo-text tokens from visual features, these methods translate multimodal composed queries into a unified text-only target description for retrieval, thereby casting CIR as a text-to-image retrieval problem under frozen VLPs. For instance, CIReVL~\cite{karthik2023vision} reformulates CIR through LLM-based target description generation, while OSrCIR~\cite{tang2025reason} directly employs a pretrained multimodal LLM as an intent reasoner with chain-of-thought prompts to enable efficient one-stage retrieval. Compared with pseudo-word learning, this line is better at capturing high-level semantic intent, but its retrieval quality depends heavily on whether the generated query can preserve the relevant visual context of the reference image.

Despite their effectiveness, existing zero-shot CIR methods~\cite{saito2023pic2word,yang2024ldre,tang2025reason} still exhibit inherent limitations in optimizing VLP-aligned multimodal composed query intent representations, particularly in fully exploiting complementary cross-modal information and correctly preserving the intended visual instance from the reference image.  
On the one hand, pseudo-word-based approaches, as shown in Figure~\ref{fig:motivation} (A), encode global- or local-level visual features into pseudo-text tokens and concatenate them with the modification text.
Although this strategy preserves reference-image information, it also injects intent-irrelevant visual noise into the composed query, such as background clutter, unrelated objects, or unchanged content.
Moreover, the resulting query often resembles a simplistic concatenation of reference content and an edit instruction, rather than a concise and target-oriented description of the intended image, making it difficult for native VLP encoders to effectively comprehend the underlying multimodal intent.
On the other hand, MLLM-based reasoning methods, as illustrated in Figure~\ref{fig:motivation} (C), can generate semantically coherent target descriptions by reasoning over the reference image and modification text.
However, they may omit explicit visual instantiation cues from the reference image, leading to ambiguity between the intended object category and the specific reference instance, especially under fine-grained modifications such as object attributes.
In other words, although these descriptions may capture the general semantic intent, they can weaken the correct grounding of the target image to the concrete visual instance in the reference image.
In addition, both paradigms may produce queries that are semantically plausible but suboptimal for retrieval: pseudo-word-based methods rely on pseudo-token concatenation, while MLLM-based methods often generate overly elaborate descriptions.
Such queries can be difficult for frozen VLP retrieval backbones, such as CLIP and BLIP, to interpret effectively, since these models typically favour concise, direct, and VLP-aligned descriptions of the target image.

\noindent\textbf{Motivation.}
To address the above issues, we are motivated to revisit the formulation of zero-shot CIR from the perspective of optimizing VLP-aligned multimodal intent representations for effective composed image retrieval.
An ideal composed query should directly describe the intended target image, rather than express the modification as an edit-style instruction, while preserving intent-relevant visual instances from the reference image and suppressing irrelevant visual noise.
However, existing paradigms struggle to simultaneously satisfy these requirements.
To this end, we combine the complementary strengths of pseudo-word-based and MLLM-based paradigms while mitigating their inherent weaknesses.
As shown in Figure~\ref{fig:motivation}(B), we guide MLLMs to generate target-oriented descriptions that explicitly depict the intended target image instead of producing change-oriented instructions.
Meanwhile, to avoid losing instance-specific information, we ground the reasoning process with correct visual instantiation from the reference image.
Rather than injecting the entire reference image representation into the query, we further disentangle fine-grained visual instance cues from global visual features and contextually integrate them with the reasoned textual query.
The resulting hybrid-modal intent representation is semantically explicit, visually grounded to the correct reference instance, and better aligned with the frozen VLP retrieval space. 

\noindent\textbf{The Proposed Method.} We propose VMIR-CVI, a zero-shot CIR framework that optimizes VLP-aligned multimodal intent representations with correct visual instantiation cues. 
First, the VLP-aligned Multimodal Intent Reasoning (VMIR) module guides MLLMs to generate target-oriented descriptions that conform to the VLP retrieval paradigm, while explicitly reasoning over intent-consistent visual instances from the reference image. Next, the Training-Free Visual Instance Disentanglement (TVID) module decouples fine-grained visual instance cues from global visual features, thereby preserving instance-specific information while suppressing intent-irrelevant noise. Finally, the Hybrid-modal Intent Alignment and Fusion (HIAF) module, built upon VMIR and TVID, leverages pre-trained VLP priors to contextually integrate the disentangled visual instance cues with the reasoned textual query, forming a coherent hybrid-modal intent representation. Together, these components reduce the semantic gap between composed queries and target images, producing concise, visually grounded, and VLP-compatible queries for precise and robust zero-shot CIR.

We summarize our contributions as follows:
 \begin{itemize} 
     \item We propose a novel VMIR-CVI, a unified zero-shot CIR paradigm that optimizes VLP-aligned multimodal intent representations by effectively integrating fine-grained visual pseudo-word embeddings with VLP-aligned MLLM-based intent reasoning, bridging multimodal intents and targets while overcoming reference inconsistency and intent-irrelevant noise. 
     \item  We introduce VMIR, an MLLM-based VLP-aligned CoT reasoning strategy that maximizes the VLP backbone’s alignment capabilities, to generate retrieval-aligned textual queries with explicit reasoned target-consistent visual references.
     \item We propose TVID, a training-free visual instance disentanglement module that decouples instance-level cues from frozen VLP visual features without additional optimization.
     \item We propose a lightweight Hybrid-modal Intent Alignment and Fusion (HIAF) module that integrates VLP-aligned textual intent and correctly instantiated visual cues into a coherent intent embedding via pretraining, enabling accurate and robust zero-shot CIR.
\end{itemize}
Extensive experiments on CIRR, CIRCO, and FashionIQ demonstrate that VMIR-CVI consistently outperforms existing zero-shot CIR methods, achieving new state-of-the-art results.

\section{Related Work}
\textbf{Zero-shot CIR with Visual Pseudo-word Learning.} For zero-shot composed image retrieval, pseudo-word learning is a mainstream approach \cite{tu2025mllm,Baldrati_2023_ICCV,Li_2025_ICCV,Tang_2025_CVPR,jang2024spherical}. Its core idea is to map the visual features of the reference image into the embedding space of a pre-trained vision-language model (such as CLIP \cite{radford2021learning}) and represent it as one or more ``pseudo-word'' tokens. This transforms the multimodal retrieval task into a standard text-to-image retrieval. Early representative methods, e.g. Pic2Word \cite{saito2023pic2word}, used a lightweight network to map images into word tokens, achieving semantic fusion and flexible composition in the language space. SEARLE\cite{Baldrati_2023_ICCV} and its improved version, iSEARLE \cite{2024iSEARLE}, introduced a two-stage process that combines optimized textual inversion with GPT-driven semantic regularization to generate pseudo-words, using knowledge distillation to enhance inference efficiency and semantic consistency. Addressing the limitation that global tokens struggle to capture fine-grained visual details, FTI4CIR \cite{lin2024fine} proposed a fine-grained textual inversion network, mapping images to topic-oriented and attribute-oriented tokens, significantly improving the modeling of complex visual attributes. Moreover, recent studies such as PrediCIR \cite{tang2025missing} further combined ``world models'' to adaptively predict missing target content in the reference image within latent space, thereby compensating for the lack of visual information before mapping it to pseudo-tokens. 
However, these methods directly learn mappings from global references, ignoring co-occurrence attentional associations between visual instances, resulting in visual noise interference in the pseudo-word embeddings. In addition, most existing methods simply concatenate pseudo-mappings with the modification text, which is not an optimal strategy for adapting composed queries to VLP backbones.

\textbf{Zero-shot CIR with LLM-guided Intent Reasoning and Understanding.}
In recent years, intent reasoning and understanding methods based on large language models (LLMs) have demonstrated significant advantages in zero-shot composed image retrieval \cite{karthik2023vision,tang2025reason,sun2025cotmr,yang2024ldre,li2024improving}. The core of these methods lies in leveraging the model’s powerful logical reasoning abilities to analyze the ``semantic increment'' between the reference image and the modification text. 
Karthik et al. propose CIReVL \cite{karthik2023vision}, a training-free zero-shot compositional image retrieval pipeline that integrates pre-trained VLMs and LLMs. It generates query image captions via VLMs, uses LLMs to recombine captions with textual modifiers into target descriptions, and retrieves images via CLIP. OSrCIR \cite{tang2025reason} directly utilizes a multimodal large model (MLLM) to process both image and text through a reflective chain-of-thought framework, effectively avoiding the loss of visual information caused by the traditional two-stage methods during image description.  However, they rely solely on reasoned textual queries while neglecting effective visual reference cues, leading to inconsistencies in intent instances between the reference and target images. In addition, existing MLLM-based intent reasoning methods rarely optimize the reasoning content when generating potential target descriptions, which leads to the inability to fully leverage the performance of VLP dependencies.

\begin{figure*}[t] 
    	\centering
    	\includegraphics[width=1\linewidth]{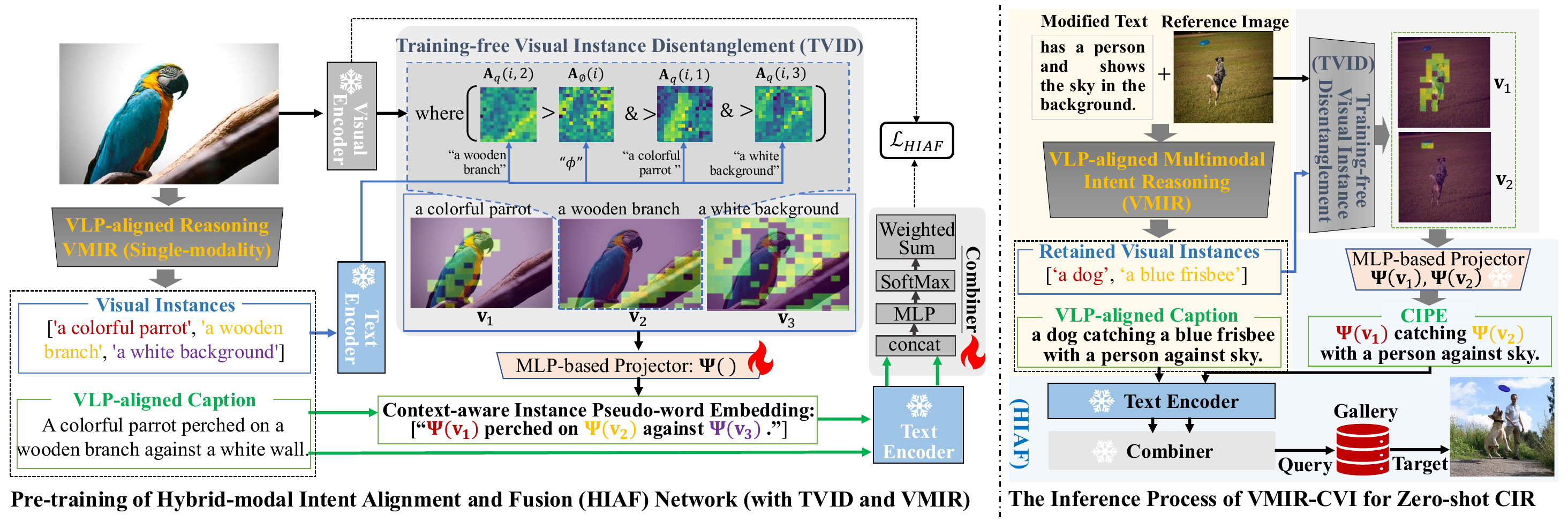}
    	\vspace{-2em}
     \caption{Illustration of our proposed VMIR-CVI framework for zero-shot CIR, including (1) the pre-training of our proposed hybrid-modal intent alignment and fusion (HIAF) with VLP-aligned intent reasoning (VMIR, here only using single-modality to obtain the image caption with instances) and training-free visual instance disentanglement (TVID), and (2) the inference process of the whole VMIR-CVI for zero-shot CIR testing (best viewed in color).
     }
    	\label{Framework}    
     	\vspace{-2em}
    \end{figure*}
\section{The Proposed Method} 
Figure~\ref{Framework} shows an overview of VMIR-CVI for zero-shot CIR. Our core innovation lies in a unified paradigm that optimizes VLP-compatible multimodal intent representations by integrating MLLM-based intent-target caption reasoning with context-aware instance pseudo-word embeddings under the frozen VLP backbone. VMIR-CVI combines the complementary strengths of two mainstream retrieval paradigms while mitigating their inherent limitations. It first generates VLP-aligned and semantically consistent target descriptions as retrieval queries (Section~\ref{VMIR}), while preserving correct fine-grained visual instantiation cues from the reference image (Section~\ref{TVID}). Finally, it reconstructs a more accurate hybrid-modal composed intent representation through Hybrid-modal Intent Alignment and Fusion pre-training (Section~\ref{HIAF}). Algorithm~\ref{alg:vmir_cvi_core} details the pre-training and inference procedures of VMIR-CVI.

\subsection{Preliminaries}

\textbf{Task Formulation.} 
Zero-shot CIR aims to retrieve a target image that satisfies a user’s composed intent, which is jointly specified by a reference image and a textual modification, without relying on paired supervision.
Formally, given a reference image $I_r$ and a modification text $T_m$, the goal is to retrieve the target image $I_t$ from a large-scale image gallery
$\mathcal{I} = \{ I_i \}_{i=1}^{N}$ ($I_r \notin \mathcal{I}$) that best matches the composed intent.
Under the zero-shot setting, no annotated triplets of the form $((I_r, T_m), I_t)$ are available during training.

Although both the query and the gallery consist of visual inputs, directly enforcing vision-to-vision alignment is suboptimal for zero-shot CIR.
The composed intent is inherently heterogeneous, typically describing instance, attribute, or relational modifications that may not be explicitly observable from the reference image alone.
In contrast, vision--language pre-trained (VLP) models, e.g., CLIP~\cite{radford2021learning}, are optimized under a large-scale text-centred alignment paradigm, where images are aligned with their textual descriptions.
Consequently, the current effective zero-shot CIR paradigm relies on constructing text-consistent query representations aligned with VLP backbones.
Let $\mathcal{F} = \{ \mathcal{F}_v, \mathcal{F}_t \}$ denote a pre-trained VLP backbone, where $\mathcal{F}_v$ and $\mathcal{F}_t$ are the vision and text encoders, respectively.
Each gallery image $I_i \in \mathcal{I}$ is encoded into a shared embedding space, as $\mathbf{v}_i = \mathcal{F}_v(I_i)$.
The core objective of zero-shot CIR is to construct and learn a better composed-intent representation model $\mathcal{G}$ that maps the multimodal query $(I_r, T_m)$ into the same VLP embedding space via $\mathcal{G}(I_r, T_m)$.

Crucially, due to the absence of annotated query--target pairs under the zero-shot setting, the composed query embedding $\mathbf{q}$ is not directly supervised to approximate the visual embedding of the target image.
Instead, it is expected to be semantically aligned with the (implicit) textual description of the desired target image.

Let $T_t$ denote the potential textual description corresponding to the target image $I_t$.
Given that modern vision--language models are trained to align images with their textual descriptions, semantic alignment between $\mathcal{F}_v(I_t)$ and $\mathcal{F}_t(T_t)$ can be reasonably assumed.
Accordingly, the learning objective of zero-shot CIR can be formulated as encouraging the composed query representation to align with the target semantic space:
\begin{equation}
\mathrm{Sim}\big( \mathcal{G}(I_r, T_m), \mathcal{F}_v(I_t) \big)
\;\approx\;
\mathrm{Sim}\big( \mathcal{G}(I_r, T_m), \mathcal{F}_t(T_t) \big).
\end{equation}
At inference time, retrieval is performed by ranking gallery images according to their similarity with the composed query embedding:
\begin{equation}
I^{*} = \arg\max_{I_i \in \mathcal{I}}
\mathrm{Sim}\big( \mathcal{G}(I_r, T_m), \mathcal{F}_v(I_i) \big),
\end{equation}
where $\mathrm{Sim}(\cdot, \cdot)$ is a similarity measure (e.g., cosine similarity) defined in the shared embedding space.

\begin{figure}[t] 
    	\centering
    	\includegraphics[width=1\linewidth]{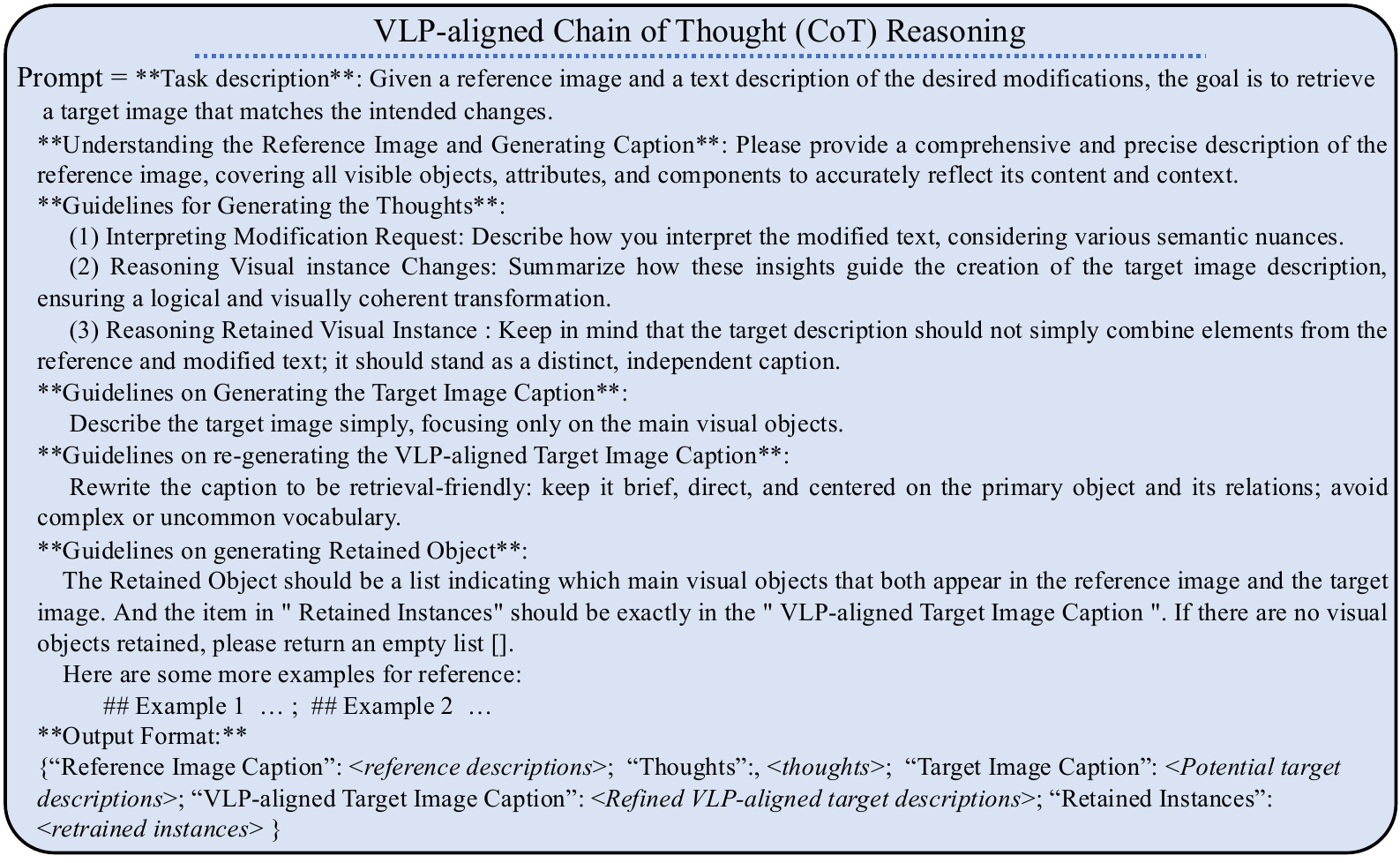}
        \caption{Multi-step CoT Prompt template for generating the VLP-aligned intent caption with few-shot VLP-query exemplars.
        }  
    	\label{fig:prompt}
    	\vspace{-1em}
\end{figure}

\noindent\textbf{Vision-Language Pre-triained Backbones.} 
Our method is built upon vision-language pre-trained (VLP) models, providing a shared embedding space for images and texts.
Following standard zero-shot CIR practice, we adopt multiple publicly available CLIP-based VLP models as backbones to validate the generality of our approach.


\subsection{VLP-aligned Multimodal Intent Reasoning} \label{VMIR}
MLLM-based intent reasoning \cite{tang2025reason,li2026stitch} is a common strategy for understanding heterogeneous and complex multimodal composed queries. However, the reasoned target descriptions may not fully exploit the alignment capability of frozen VLP backbones, since they often deviate from the native textual preference of pre-trained VLP models, i.e., concise semantic descriptions with direct logical expressions. To bridge this representation gap, we introduce the VMIR module. Specifically, VMIR incorporates a multi-step Chain-of-Thought (CoT) prompting strategy, supplemented with few-shot VLP-native query exemplars, to guide the MLLM in generating a VLP-aligned intent description $T_t$. In addition to describing the potential target content, the MLLM explicitly infers which visual instances from the reference image should be retained in the target image, thereby enabling correct visual instantiation while reducing subsequent computation. The reasoning process can be formulated as:
\begin{equation}
            (T_t, \mathcal{O}_t)={\rm MLLM}(I_r,T_m, CoT\text{-}Prompt),
\end{equation}
where the VLP-aligned $CoT\text{-}Prompt$ is shown in Figure \ref{fig:prompt} and $O_t=\{o_i\}$ will be used for subsequent multimodal fusion.

\subsection{Training-free Visual Instance Disentanglement} \label{TVID}
Unlike textual modifications, selecting correct visual cues from the reference image in zero-shot CIR is more challenging, as naive visual pseudo-word learning methods \cite{saito2023pic2word,gu2024language} usually introduce irrelevant visual noise. This is because they rely only on global coarse-grained attention-aware mapping, rather than accurately locating individual instances to eliminate interference. To address this issue, we propose a TVID module, leveraging the pre-trained visual encoder of VLP backbones to decouple instance-level features without additional training. It mainly consists of three steps.

\textbf{(1) Dense visual feature extraction.} Given a reference image $I_r$, we first extract patch-level hidden states $h_{L-1}$ from the penultimate ViT layer, which retains local instance information and avoids the over-smoothing caused by the final layer's global [CLS] attention. Then, it is forwarded through the final layer without attention computation (AttnFreeLayer), followed by a LayerNorm and visual projection (VP) operations, to obtain final dense representation $\mathcal{H}$.
\begin{equation}
\mathcal{H} = \text{VP}\big(\text{LayerNorm}(\text{AttnFreeLayer}({\rm ViT}_{L-1}(I_r)))\big),
\end{equation}
where $\mathcal{H} \in \mathbb{R}^{N \times D}$ ($N$ is the number of image patches excluding the [CLS] token, and $D$ is the feature dimension).

\textbf{(2) Instance-aware attention computation.}  
Let $\{o_1, \dots, o_K\}$ denote a set of instance keywords, which can be reasoned from a composed query by MLLMs. We encode them using the pre-trained text encoder $\mathcal{F}_t$ to obtain normalized text embeddings $\mathbf{T_o}$, as:
\begin{equation}
\mathbf{T_o} = [\mathcal{F}_t(o_1), \dots, \mathcal{F}_t(o_K)] = [\mathbf{t}_o^1, \dots, \mathbf{t}_o^K],\quad  \mathbf{t}_o^j \in \mathbb{R}^{1 \times D}.
\end{equation}

To discover visually and semantically irrelevant but highly responsive noisy patch features, we encode an empty text cue (its embedding denoted as $\mathbf{t}_{\varnothing}$=$\mathcal{F}_t(\varnothing)$) as a non-semantic text reference, shown in Figure \ref{Framework}. 
Then, similarity scores of visual patch and textual instance keywords can be computed via inner products in the shared embedding space, as:
\begin{equation}
\mathbf{A}_o = \mathcal{H} \mathbf{T}_o^\top \in \mathbb{R}^{N \times K},
 \quad
\mathbf{A}_{\varnothing} = \mathcal{H} \mathbf{t}_{\varnothing}^\top \in \mathbb{R}^{N \times 1},
\end{equation}
where $\mathbf{A}_o[i,j]$ measures the response of $i$-th patch to instance keyword $o_j$, and $\mathbf{A}_{\varnothing}[i]$ denotes the response of $i$-th patch to the non-semantic reference.

\textbf{(3) Instance-level patch disentanglement.} 
For each instance keyword $o_j$, we select patches that are uniquely activated by this keyword while suppressing cross-keyword interference and background noise.
The valid patch set $\mathcal{P}_j$ is defined as:
\begin{equation}
\mathcal{P}_j =
\left\{
i \;\middle|\;
\mathbf{A}_o[i,j] > \mathbf{A}_o[i,k],\ \forall k \neq j,
\ \text{and}\
\mathbf{A}_o[i,j] > \mathbf{A}_{\varnothing}[i]
\right\}.
\end{equation}
This strategy ensures that each retained patch responds more strongly to its corresponding instance keyword than to other keywords or the non-semantic reference, effectively disentangling instance-level visual cues.
For each instance keyword $o_j$, the selected patches are aggregated into an individual instance-level visual representation via attention-weighted pooling.
The weight assigned to patch $i \in \mathcal{P}_j$ is computed as:
\begin{equation}
w_i^{(j)} = 
\frac{\mathbf{A}_o[i,j]}{\sum_{i \in \mathcal{P}_j} \mathbf{A}_o[i,j]},
\end{equation}
and the corresponding visual instance embedding is obtained as:
\begin{equation}
\mathbf{v}_j = \sum_{i \in \mathcal{P}_j} w_i^{(j)} \, \mathcal{H}_i \in \mathbb{R}^{D}.
\end{equation}

Repeating this process for all instance keywords yields a set of disentangled and noise-free visual instance features $\{\mathbf{v}_1, \dots, \mathbf{v}_K\}$, which are subsequently used for hybrid intent representation.

\begin{algorithm}[t]
\caption{Pseudo code of VMIR-CVI.}
\label{alg:vmir_cvi_core}
\SetAlgoLined
\tcp*[h]{\textcolor{blue}{HIAF Pre-training Stage with VMIR and TVID Modules}} \\
\KwIn{Training set $\mathcal{N}=\{I_i\}_{i=1}^\Omega$, pre-trained  CLIP-ViT $\mathcal{F}_v$, CLIP-Text $\mathcal{F}_t$ and MLLM, batch size $B$, epochs $E$}
\KwOut{Trained Context-aware Projector $\Psi$ and $Combiner$}

Initialize Projector $\Psi$, $Combiner$ parameters\;

\linespread{0.5}\selectfont

\For{$epoch = 1$ \KwTo $E$}{
    \For{each batch}{
        \For{each \text{image} $I_i$  }{
            \textcolor{gray}{{// Reason image caption with corresponding visual instances}} \\
            $\{T_i,O_i\}$ $\gets$ $VMIR(I_i)$; \# here wtih only image input \\
            Extract image feature: $z_v$ = $\mathcal{F}_v(I_i)$\;
            Extract text feature:  $z_t$ = $\mathcal{F}_t(T_i)$\;
            \textcolor{gray}{{// Disentangle visual instance embeddings}} \\
             $\{v_1,v_2,...\}$ $\gets$ $TVID(\mathcal{F}_v|I_i,O_i)$\;
            \textcolor{gray}{{// Get context-aware intent with visual instance pseudo token}} \\
            $\tilde{z}_t$ $\gets$ $CIPE(\tilde{T}_i = \{ w_1, \ldots,(w_m\rightarrow\Psi(v_k)),w_M \})$\;
            \textcolor{gray}{{// Hybrid-modal intent representation fusion}} \\
            $z_h$ $\gets$ $Combiner(z_t,\tilde{z}_t)$;
        }
        Optimized $\Psi$ and $Combiner$ by $\mathcal{L}_{HIAF}$($z_h,z_v$) in Eq.\ref{hiaf}.\
    }
}
\tcp*[h]{\textcolor{blue}{Inference Process of Zero-shot Multimodal Composed Image Retrieval}} \\
\KwIn{Composed query $(I_r, T_m)$, Gallery $\mathcal{I}$, Pre-trained CLIP-ViT $\mathcal{F}_v$ and CLIP-Text $\mathcal{F}_t$ , MLLM, Trained $\Psi$ and $Combiner$}
\KwOut{Target image $I_t$}

 $(T_i,O_i)$$\gets$$VMIR(I_r,T_m)$;  \textcolor{gray}{{//{ Get VLP-aligned target caption and retained refernece instances}}} \\
 $\{v_k\}_{k=1}^K$$\gets$ $TVID(\mathcal{F}_v|I_r,O_i)$;  \textcolor{gray}{{//{ Disentangle visual reference instances}}} \\
$\tilde{z}_t$ $\gets$ $CIPE(\Psi,\{v_k\}_{k=1}^K, T_i )$; \textcolor{gray}{{// Get comtext-aware hybird intent representation}} \\
$z_h$ $\gets$ $Combiner( \mathcal{F}_t(T_i),\tilde{z}_t)$; \textcolor{gray}{{// Get hybrid-modal fused intent representation}} \\
$I_t = \arg\max_{I_i \in \mathcal{I}} \mathrm{Sim}\big( z_h, \mathcal{F}_v(I_i) \big)$. \textcolor{gray}{{// Similarity Ranking}}
\end{algorithm}

\subsection{Hybrid-modal Intent Alignment and Fusion} \label{HIAF}
While VMIR produces VLP-aligned potential target descriptions and TVID provides disentangled, noise-free visual instance representations, a critical challenge remains: \textit{how to coherently integrate heterogeneous intent cues from different modalities into a unified representation that is fully aligned with the VLP retrieval space.} Existing pseudo-word learning methods typically rely on global coarse-grained visual pseudo-tokens, whereas MLLM-based pure text reasoning neglects explicit visual cues, leading to reference inconsistency and modality imbalance.

To address this issue, we propose Hybrid-modal Intent Alignment and Fusion (HIAF), a pre-training strategy that unifies reasoned VLP-aligned textual intent and disentangled visual instance cues into a unified hybrid-modal intent representation. Specifically, it is used to (1) embed correct fine-grained visual instances into VLP-aligned target descriptions in a context-aware manner, and (2) adaptively fuse textual and visual-driven intent signals via a learnable \textit{Combiner}, ensuring compatibility with pre-trained VLP encoders.

\noindent\textbf{Pre-training: Hybrid-modal Intent Alignment and Fusion Learning.}
The goal of the pre-training stage is to learn a robust hybrid-modal intent representation that is well aligned with the VLP retrieval space. This stage consists of three components: hybrid-modal intent construction, combiner-based hybrid-modal fusion, and contrastive pre-training objectives.

\textbf{(1) Hybrid-modal intent construction.} 
Given an image $I$, which serves as both the reference and the target during the pre-training stage, VMIR (single visual input) generates a textual description $T=\{w_1,\ldots,w_M\}$ that conforms to the native textual query format of the VLP and faithfully describes the image content. Meanwhile, VMIR identifies a set of salient visual instance keywords present in the image, i.e.,
$\mathcal{O} = \{ o_k \}_{k=1}^K$.

In parallel, TVID disentangle the corresponding fine-grained visual instance representations $\{ v_1, \ldots, v_K \}$ ($v_k \in \mathbb{R}^D$) from the pretrained visual encoder $\mathcal{F}_v$ of the same image.

To inject instance-level visual information into the text intent space, we propose a \textbf{context-aware instance pseudo word embedding strategy (CIPE)} in Figure~\ref{Framework}. This strategy, based on the visual instance decoupling of TVID, further replaces the corresponding context keyword positions of the VLP-aligned description with the visual instance embeddings. This effectively solves the problems of global visual noise and simple splicing of hybrid modalities in traditional global vision pseudo-word learning methods \cite{saito2023pic2word,lin2024fine}. Specifically, we also employ an MLP-based projection function $\Psi(\cdot)$, which maps each visual instance embedding to the VLP textual embedding space, yielding instance pseudo-word embeddings $p_k = \Psi(v_k)$.
We then construct a hybrid-modal ``textual'' sequence by replacing the corresponding visual instance keywords in VLP-aligned caption $T$ with their pseudo-word embeddings:
\begin{equation}
\tilde{T} = \{ w_1, \ldots,(w_m\rightarrow p_k),w_M \}, 
\end{equation}
where the $m$-th instance keyword $w$ is replaced by $k$-th instance pseudo-word $p_k$ ($k\in [1,K]$) with the same semantics and the others are retained from $T$.
This hybrid-modal sequence preserves the linguistic structure of the original description while explicitly grounding semantic tokens with visual instance information.

\textbf{(2) Hybrid-modal fusion via combiner.} 
Although the hybrid-modal sequence $\tilde{T}$ encodes visual instance cues, relying solely on it may bias the representation toward local visual details. Conversely, the original textual description $T$ provides strong explicit global semantics but lacks visual instance-level grounding. To balance these complementary properties, we introduce a learnable \textit{Combiner} to fuse the two representations.

Specifically, both types of intent descriptions are encoded using a shared VLP text encoder $\mathcal{F}_t(\cdot)$:
\begin{equation}
z_t = \mathcal{F}_t(T), \quad \tilde{z}_t = \mathcal{F}_t(\tilde{T}).
\end{equation}

As shown in Figure \ref{Framework}, the Combiner $\mathcal{C}()$ dynamically integrates these embeddings to produce the final hybrid-modal intent representation $z_h$, where the Combiner is implemented as a lightweight MLP with a soft weighting mechanism. It first projects and concatenates them $[z_t:\tilde{z}_t]$ to form a joint representation, and then estimates the fusion weights via an MLP followed by a softmax operation: $\{\beta_1, \beta_2\} = \mathrm{SoftMax}(\mathrm{MLP}([z_t:\tilde{z}_t]))$
Finally, the hybrid-modal intent representation is obtained through a weighted aggregation, allowing the model to adaptively balance global semantic intent and instance-level visual grounding across different training samples. 
\begin{equation}
z_h = \mathcal{C}(z_t, \tilde{z}_t)=\beta_1 z_t+ \beta_2 \tilde{z}_t.
\end{equation}
Through pre-training, the Combiner learns a stable fusion strategy that is compatible with the VLP embedding space.

\textbf{(3) Pre-training objective.}
To align hybrid-modal intent representation $z_h$ with corresponding image embedding, we adopt a CLIP-style bidirectional (text-to-image and image-to-text) contrastive learning objective \cite{radford2021learning}. The overall pre-training objective is:
\begin{equation} \label{hiaf}
\mathcal{L}_{HIAF} = - \frac{1}{|\mathcal{B}|} \sum_{i \in \mathcal{B}}(\log 
\frac{\exp(\langle z_h^i, z_v^i \rangle / \tau)}
{\sum_j \exp(\langle z_h^i, z_v^j \rangle / \tau)}) +
 \log 
\frac{\exp(\langle z_v^i, z_h^i \rangle / \tau)}
{\sum_j \exp(\langle z_v^i, z_h^j \rangle / \tau)}) ,
\end{equation}
where $z_v^i$ is the i-th image embedding from $\mathcal{F}_v$ and  $\tau$ is the temperature coefficient.

\noindent\textbf{Zero-shot CIR Inference.}
At inference, VMIR is applied to the composition query to generate a VLP-aligned target description $T_t$. Following the same procedure as in pre-training, the reasoned correct reference instance keywords in $T_t$ are replaced with the corresponding instance pseudo-text tokens $\mathbf{p}_k$, constructing the hybrid-modal sequence $\tilde{T}_t$:
\begin{equation}
\tilde{T}_t = \{ \tilde{w}_1, \dots, \tilde{w}_M \}, \quad \tilde{w}_i\in \{w_i\ or\ p_k\}.
\end{equation}

The hybrid-modal sequence $\tilde{T}_t$ is encoded via the shared text encoder $\mathcal{F}_t(\cdot)$, and fused with the pure textual representation using the Combiner $\mathcal{C}(\cdot)$ to produce the final hybrid intent representation:
\begin{equation}
\mathbf{z}_h = \mathcal{C}(\mathcal{F}_t(T_t), \mathcal{F}_t(\tilde{T}_t)).
\end{equation}

Zero-shot composed image retrieval is then performed by computing the similarity between $\mathbf{z}_h$ and candidate image embeddings $\mathbf{z}_v = \mathcal{F}_v(I)$, ensuring full compatibility with the pre-trained hybrid-modal alignment space.

\section{Experimental Setup}
\textbf{Dataset.} 
To verify the effectiveness, we conduct extensive experiments on three standard composed image retrieval benchmarks, \textit{i.e.,} CIRR~\cite{liu2021image}, FashionIQ \cite{wu2021fashion} and CIRCO \cite{Baldrati_2023_ICCV}.
CIRR is an open-world scene dataset with 36,554 triplets from 21,552 NLVR2 images~\cite{suhr2018corpus}, split 8:1:1 for training, validation, and testing. Fashion-IQ contains 30,134 triplets from 77,684 fashion images, divided into Dress, Shirt, and Toptee categories. The CIRCO dataset is an open-domain benchmark using images from COCO image \cite{2014Microsoft} . It includes 1,020 query triplets, split into 220 validation and 800 test triplets. Fashion-IQ targets a specialised domain, whereas CIRR and CIRCO emphasise diverse object interactions in real-world scenes. Evaluations on both datasets better reflect the method’s real-world generalization ability.

\noindent\textbf{Evaluation metrics.}
Following prior work \cite{gu2024language}, we evaluate our model across three benchmarks. For CIRR, we report Recall@K (K=1, 5, 10, 50). For CIRCO, we utilize mAP@K (K=5, 10, 25, 50) \cite{2020A} to evaluate retrieval performance. For FashionIQ, we report Recall@10 and Recall@50 for each of the three categories, along with their mean values, to assess the model's generalization capability within specific fashion domains. 

\noindent\textbf{Implementation Details.}
During the pre-training of HIAF, we utilized a subset of ImageNet2012~\cite{deng2009imagenet} containing 100,000 images as the image corpus. The Qwen3-VL-30B-A3B-Instruct~\cite{Qwen3-VL} model was employed to generate the corresponding image captions and instance annotations. For training the Projector $\Psi$ and the Combiner $\mathcal{C}$, we used a learning rate of $2\times10^{-5}$ and a batch size of 32. The temperature parameter in Eq.~\eqref{hiaf} was set to $\tau=0.01$. Training was conducted for 3 epochs on a single NVIDIA RTX2080Ti GPU, taking approximately 3 hours and using around 10 GB of GPU memory. To ensure reproducibility, we fixed the random seed to 43 in all experiments. During inference, we used Qwen3-VL-30B-A3B-Instruct to perform chain-of-thought reasoning and generate the target caption together with the retained instances.

\section{Experiments}
In this section, we report experimental results and provide detailed analyses to address the following research questions:

\noindent\textbf{--RQ1:} How does the performance of the proposed VMIR-CVI for the zero-shot CIR task compare to existing SOTA methods?

\noindent\textbf{--RQ2:} How do the components of the proposed VMIR-CVI affect the zero-shot retrieval performance? And how do the proposed VMIR, TVID, and HIAF perform compared to other implementations? 


\noindent\textbf{--RQ3:} How does the inference efficiency of VMIR-CVI? 

\noindent\textbf{--RQ4:} How do VMIR and TVID improve qualitative results of multimodal intent reasoning and fine-grained instance disentanglement?

\begin{table*}[t]
\caption{Performance comparison on CIRR under the open-world setting on the test set. Boldface and underlining are used to indicate the best and second-best scores, respectively. ``*'' denotes the p-value<0.05 over the second-best baseline.}
\vspace{-1em}
\renewcommand\tabcolsep{3.6pt}
\label{tab:compare_cirr}
\begin{tabular}{cc|ccccccccc}
\hline
\multicolumn{1}{c|}{\multirow{2}{*}{Arch}} & \multirow{2}{*}{Method} & \multicolumn{5}{c|}{Recall@k} & \multicolumn{4}{c}{Recall$_{Subset@k}$} \\
\multicolumn{1}{c|}{} & & k=1 & k=5 & k=10 & k=50 & \multicolumn{1}{c|}{Avg.} & k=1 & k=2 & k=3 & Avg. \\ \hline

\multicolumn{1}{c|}{\multirow{5}{*}{ViT-B/32}} & SEARLE \cite{Baldrati_2023_ICCV} $_{(ICCV'2023)}$ & 24.00 & 53.42 & 66.82 & 89.78 & \multicolumn{1}{|c|}{58.51} & 54.89 & 76.60 & 88.19 & \multicolumn{1}{|c}{73.23} \\
\multicolumn{1}{c|}{} & CIReVL \cite{karthik2023vision} $_{(ICLR'2024)}$ & 23.94 & 52.51 & 66.00 & 86.95 & \multicolumn{1}{|c|}{57.35} & 60.17 & 80.05 & 90.19 & \multicolumn{1}{|c}{76.80} \\
\multicolumn{1}{c|}{} & LDRE \cite{yang2024ldre} $_{(SIGIR'2024)}$ & {\ul 25.69} & {\ul 55.13} & {\ul 69.04} & {\ul 89.90} & \multicolumn{1}{|c|}{{\ul 59.94}} & 60.53 & 80.65 & 90.70 & \multicolumn{1}{|c}{77.29} \\
\multicolumn{1}{c|}{} & OSrCIR \cite{tang2025reason} $_{(CVPR'2025)}$ & 25.42 & 54.54 & 68.19 & - & \multicolumn{1}{|c|}{-} & {\ul 62.31} & {\ul 80.86} & {\ul 91.13} & \multicolumn{1}{|c}{{\ul 78.10}} \\
\multicolumn{1}{c|}{} & \textbf{VMIR-CVI $_{(Ours)}$} & \textbf{29.59*} & \textbf{60.34*} & \textbf{72.41*} & \textbf{91.01} & \multicolumn{1}{|c|}{\textbf{63.34}} & \textbf{65.71*} & \textbf{83.66*} & \textbf{92.84*} & \multicolumn{1}{|c}{\textbf{80.74}} \\ \hline

\multicolumn{1}{c|}{\multirow{12}{*}{ViT-L/14}} & Pic2Word \cite{saito2023pic2word} $_{(CVPR'2023)}$ & 23.90 & 51.70 & 65.30 & 87.80 & \multicolumn{1}{|c|}{57.18} & - & - & - & \multicolumn{1}{|c}{-} \\
\multicolumn{1}{c|}{} & SEARLE \cite{Baldrati_2023_ICCV} $_{(ICCV'2023)}$ & 24.24 & 52.48 & 66.29 & 88.58 & \multicolumn{1}{|c|}{57.90} & 53.76 & 75.01 & 88.19 & \multicolumn{1}{|c}{72.32} \\
\multicolumn{1}{c|}{} & LinCIR \cite{gu2024language} $_{(CVPR'2024)}$ & 25.04 & 53.25 & 66.68 & - & \multicolumn{1}{|c|}{-} & 57.11 & 77.37 & 88.89 & \multicolumn{1}{|c}{74.46} \\
\multicolumn{1}{c|}{} & Context-I2W \cite{tang2024context} $_{(AAAI'2024)}$ & 25.60 & 55.10 & 68.50 & 89.80 & \multicolumn{1}{|c|}{59.75} & - & - & - & \multicolumn{1}{|c}{-} \\
\multicolumn{1}{c|}{} & CIReVL \cite{karthik2023vision} $_{(ICLR'2024)}$ & 24.55 & 52.31 & 64.92 & 86.34 & \multicolumn{1}{|c|}{57.03} & 59.54 & 79.88 & 89.69 & \multicolumn{1}{|c}{76.37} \\
\multicolumn{1}{c|}{} & LDRE \cite{yang2024ldre} $_{(SIGIR'2024)}$ & 26.53 & 55.57 & 67.54 & 88.50 & \multicolumn{1}{|c|}{59.54} & 60.43 & 80.31 & 89.90 & \multicolumn{1}{|c}{76.88} \\
\multicolumn{1}{c|}{} & MOA \cite{li2025rethinking} $_{(SIGIR'2025)}$ & 27.10 & 56.50 & 69.20 & \underline{90.00} & \multicolumn{1}{|c|}{60.70} & - & - & - & \multicolumn{1}{|c}{-} \\
\multicolumn{1}{c|}{} & MLLM-I2W \cite{bao2025mllm} $_{(COLING'2025)}$ & 28.30 & 57.90 & 70.20 & \textbf{93.90} & \multicolumn{1}{|c|}{{\ul 62.58}} & - & - & - & \multicolumn{1}{|c}{-} \\
\multicolumn{1}{c|}{} & PrediCIR \cite{tang2025missing} $_{(CVPR'2025)}$ & 27.20 & 57.00 & 70.20 & - & \multicolumn{1}{|c|}{-} & - & - & - & \multicolumn{1}{|c}{-} \\
\multicolumn{1}{c|}{} & OSrCIR \cite{tang2025reason} $_{(CVPR'2025)}$ & {\ul 29.45} & {\ul 57.68} & {\ul 69.86} & - & \multicolumn{1}{|c|}{-} & {\ul 62.12} & {\ul 81.92} & {\ul 91.10} & \multicolumn{1}{|c}{{\ul 78.38}} \\
\multicolumn{1}{c|}{} & \textbf{VMIR-CVI $_{(Ours)}$} & \textbf{30.00*} & \textbf{60.22*} & \textbf{70.82*} & 89.71 & \multicolumn{1}{|c|}{\textbf{62.69}} & \textbf{65.93*} & \textbf{84.17*} & \textbf{92.70*} & \multicolumn{1}{|c}{\textbf{80.93}} \\ \hline

\multicolumn{1}{c|}{\multirow{6}{*}{ViT-G/14}} & LinCIR \cite{gu2024language} $_{(CVPR'2024)}$ & 35.25 & 64.72 & 76.05 & - & \multicolumn{1}{|c|}{-} & 63.35 & 82.22 & 91.98 & \multicolumn{1}{|c}{79.18} \\
\multicolumn{1}{c|}{} & CIReVL \cite{karthik2023vision} $_{(ICLR'2024)}$ & 34.65 & 64.29 & 75.06 & 91.66 & \multicolumn{1}{|c|}{66.42} & 67.95 & 84.87 & 93.21 & \multicolumn{1}{|c}{82.01} \\
\multicolumn{1}{c|}{} & LDRE \cite{yang2024ldre} $_{(SIGIR'2024)}$ & 36.15 & 66.39 & 77.25 & {\ul 93.95} & \multicolumn{1}{|c|}{{\ul 68.44}} & 68.82 & {\ul 85.66} & {\ul 93.76} & \multicolumn{1}{|c}{{\ul 82.75}} \\
\multicolumn{1}{c|}{} & PrediCIR \cite{tang2025missing} $_{(CVPR'2025)}$ & 37.00 & 66.10 & {\ul 77.90} & - & \multicolumn{1}{|c|}{-} & - & - & - & \multicolumn{1}{|c}{-} \\
\multicolumn{1}{c|}{} & OSrCIR \cite{tang2025reason} $_{(CVPR'2025)}$ & {\ul 37.26} & {\ul 67.25} & {77.33} & - & \multicolumn{1}{|c|}{-} & {\ul 69.22} & 85.28 & 93.55 & \multicolumn{1}{|c}{82.68} \\
\multicolumn{1}{c|}{} & \textbf{VMIR-CVI $_{(Ours)}$} & \textbf{37.81*} & \textbf{70.41*} & \textbf{80.43*} & \textbf{93.95} & \multicolumn{1}{|c|}{\textbf{70.65}} & \textbf{73.52*} & \textbf{88.29*} & \textbf{94.58*} & \multicolumn{1}{|c}{\textbf{85.46}} \\ \hline
\end{tabular}
\end{table*}

All comparisons are conducted under the zero-shot CIR setting. Unless otherwise specified, the evaluation follows the dataset splits, metrics, and implementation settings described in Section~4. For CIRCO and CIRR, we report results on the test set; for FashionIQ, we report validation-set results following prior work. We compare VMIR-CVI with two representative families of zero-shot CIR methods: visual pseudo-word learning methods such as Pic2Word~\cite{saito2023pic2word}, SEARLE~\cite{Baldrati_2023_ICCV}, LinCIR~\cite{gu2024language}, Context-I2W~\cite{tang2024context}, and LDRE~\cite{yang2024ldre}; and MLLM/LLM-based reasoning methods such as CIReVL~\cite{karthik2023vision}, OSrCIR~\cite{tang2025reason}, and related recent variants. This comparison protocol allows us to examine whether VMIR-CVI benefits from both accurate textual intent reasoning and explicit visual instantiation.

\subsection{State-of-the-art Comparisons (RQ1)}

\subsubsection{Overall comparison protocol}

To answer RQ1, we compare VMIR-CVI with previous zero-shot CIR baselines on three benchmarks: CIRCO, CIRR, and FashionIQ. These datasets differ substantially in retrieval granularity and domain characteristics. CIRCO evaluates open-domain composed retrieval with multiple semantically valid targets and is measured by mAP@K. CIRR focuses on open-world natural images and reports both global Recall@K and subset Recall@K, where the subset metric is particularly sensitive to fine-grained disambiguation among visually similar candidates. FashionIQ is a fashion-domain benchmark and evaluates category-specific attribute modifications using Recall@10 and Recall@50.

We also evaluate three CLIP-based VLP backbones, namely ViT-B/32, ViT-L/14, and ViT-G/14, when corresponding baseline results are available. This is important because a zero-shot CIR method should not only work under one backbone size. If the proposed query representation is genuinely aligned with the VLP retrieval space, its advantage should remain visible when the visual-language backbone changes.

\subsubsection{Results on CIRR}
Table~\ref{tab:compare_cirr} reports the results on CIRR under the open-world setting. VMIR-CVI consistently delivers the strongest overall performance across all three VLP backbones, achieving the best average Recall@k and average Recall$_{\mathrm{Subset}}$@k in each backbone group. Moreover, most improvements at small retrieval cutoffs and on the subset metrics are statistically significant, demonstrating the effectiveness and robustness of the proposed VLP-aligned multimodal intent representation.

With ViT-B/32, VMIR-CVI improves the average Recall@k by 3.40 points over the strongest complete baseline, while increasing the average Recall$_{\mathrm{Subset}}$@k by 2.64 points. The improvements are particularly evident at Recall@1 and Recall@5, indicating that VMIR-CVI more reliably ranks the intended target at the top of the retrieval list. With ViT-L/14, VMIR-CVI achieves the best Recall@1, Recall@5, and Recall@10, as well as the highest average Recall@k. Although MLLM-I2W obtains a higher Recall@50, the stronger performance of VMIR-CVI at smaller retrieval cutoffs indicates more precise target identification rather than merely broader retrieval coverage. In addition, VMIR-CVI improves the average subset recall by 2.55 points over OSrCIR. With ViT-G/14, VMIR-CVI further improves the average Recall@k and average Recall$_{\mathrm{Subset}}$@k by 2.21 and 2.71 points, respectively, while matching the best Recall@50 result. The consistent improvements from ViT-B/32 to ViT-G/14 demonstrate that the benefits of VMIR-CVI are preserved as the capacity of the underlying VLP backbone increases.

These results directly support the motivation behind the proposed framework. VMIR reformulates the composed query as a concise and target-oriented description, rather than retaining an edit-style instruction. By explicitly reasoning about which visual instances and attributes should be preserved or modified, VMIR generates textual queries that better conform to the native text space and retrieval paradigm of frozen VLPs. This reduces the semantic mismatch between the composed query and the target image, contributing to the consistent improvements at the top of the ranked list.

The gains on the subset evaluation further validate the importance of correct visual instantiation. CIRR subset retrieval requires the model to distinguish the target from a small group of semantically and visually similar candidates, making it particularly sensitive to instance-level ambiguity and subtle attribute changes. TVID addresses this challenge by disentangling fine-grained, intent-consistent visual instance cues from global visual features while suppressing irrelevant objects and background content. HIAF then contextually integrates these correctly instantiated visual cues with the VLP-aligned textual intent produced by VMIR. Consequently, VMIR-CVI forms a coherent hybrid-modal representation that is both semantically explicit and visually grounded.

Overall, the stable improvements across standard and subset retrieval demonstrate that VMIR-CVI effectively combines the complementary strengths of MLLM-based intent reasoning and visual pseudo-word representations. 

\begin{table*}[t]
\caption{Performance comparison on CIRCO under the open-world setting on the test set. Boldface and underlining are used to indicate the best and second-best scores, respectively. ``*'' denotes the p-value<0.05 over the second-best baseline.}
\vspace{-1em}
\renewcommand\tabcolsep{10pt}
\label{tab:compare_circo}
\begin{tabular}{cc|cccc|c}
\hline
\multicolumn{1}{c|}{\multirow{2}{*}{Arch}} 
& \multirow{2}{*}{Method} 
& \multicolumn{4}{c|}{mAP@k} 
& \multirow{2}{*}{Avg.} \\
\multicolumn{1}{c|}{} 
&  
& k=5 
& k=10 
& k=25 
& k=50 
& \\ \hline

\multicolumn{1}{c|}{\multirow{5}{*}{ViT-B/32}} 
& SEARLE \cite{Baldrati_2023_ICCV} $_{(ICCV'2023)}$ 
& 9.35 & 9.94 & 11.13 & 11.81 & 10.56 \\

\multicolumn{1}{c|}{} 
& CIReVL \cite{karthik2023vision} $_{(ICLR'2024)}$ 
& 14.94 & 15.42 & 17.00 & 17.82 & 16.30 \\

\multicolumn{1}{c|}{} 
& LDRE \cite{yang2024ldre} $_{(SIGIR'2024)}$ 
& 17.96 & 18.32 & 20.21 & 21.11 & 19.40 \\

\multicolumn{1}{c|}{} 
& OSrCIR \cite{tang2025reason} $_{(CVPR'2025)}$ 
& {\ul 18.04} & {\ul 19.17} & {\ul 20.94} & {\ul 21.85} 
& {\ul 20.00} \\

\multicolumn{1}{c|}{} 
& \textbf{VMIR-CVI $_{(Ours)}$} 
& \textbf{19.79*} & \textbf{20.68*} & \textbf{22.84*} 
& \textbf{23.88*} & \textbf{21.80*} \\ \hline

\multicolumn{1}{c|}{\multirow{10}{*}{ViT-L/14}} 
& Pic2Word \cite{saito2023pic2word} $_{(CVPR'2023)}$ 
& 8.72 & 9.51 & 10.64 & 11.29 & 10.04 \\

\multicolumn{1}{c|}{} 
& SEARLE \cite{Baldrati_2023_ICCV} $_{(ICCV'2023)}$ 
& 11.68 & 12.73 & 14.33 & 15.12 & 13.47 \\

\multicolumn{1}{c|}{} 
& LinCIR \cite{gu2024language} $_{(CVPR'2024)}$ 
& 12.59 & 13.58 & 15.00 & 15.85 & 14.26 \\

\multicolumn{1}{c|}{} 
& Context-I2W \cite{tang2024context} $_{(AAAI'2024)}$ 
& 13.04 & 14.62 & 16.14 & 17.16 & 15.24 \\

\multicolumn{1}{c|}{} 
& CIReVL \cite{karthik2023vision} $_{(ICLR'2024)}$ 
& 18.57 & 19.01 & 20.89 & 21.80 & 20.07 \\

\multicolumn{1}{c|}{} 
& LDRE \cite{yang2024ldre} $_{(SIGIR'2024)}$ 
& 23.35 & 24.03 & 26.44 & 27.50 & 25.33 \\

\multicolumn{1}{c|}{} 
& MOA \cite{li2025rethinking} $_{(SIGIR'2025)}$ 
& 15.30 & 17.10 & 18.50 & 19.30 & 17.55 \\

\multicolumn{1}{c|}{} 
& PrediCIR \cite{tang2025missing} $_{(CVPR'2025)}$ 
& 15.70 & 17.10 & 18.60 & 19.30 & 17.68 \\

\multicolumn{1}{c|}{} 
& OSrCIR \cite{tang2025reason} $_{(CVPR'2025)}$ 
& {\ul 23.87} & {\ul 25.33} & {\ul 27.84} & {\ul 28.97} 
& {\ul 26.50} \\

\multicolumn{1}{c|}{} 
& \textbf{VMIR-CVI $_{(Ours)}$} 
& \textbf{26.48*} & \textbf{27.72*} & \textbf{30.26*} 
& \textbf{31.36*} & \textbf{28.96*} \\ \hline

\multicolumn{1}{c|}{\multirow{6}{*}{ViT-G/14}} 
& LinCIR \cite{gu2024language} $_{(CVPR'2024)}$ 
& 19.71 & 21.01 & 23.13 & 24.18 & 22.01 \\

\multicolumn{1}{c|}{} 
& CIReVL \cite{karthik2023vision} $_{(ICLR'2024)}$ 
& 26.77 & 27.59 & 29.96 & 31.03 & 28.84 \\

\multicolumn{1}{c|}{} 
& LDRE \cite{yang2024ldre} $_{(SIGIR'2024)}$ 
& {\ul 31.12} & {\ul 32.24} & 34.95 & 36.03 
& {\ul 33.59} \\

\multicolumn{1}{c|}{} 
& PrediCIR \cite{tang2025missing} $_{(CVPR'2025)}$ 
& 23.70 & 24.60 & 25.40 & 26.00 & 24.93 \\

\multicolumn{1}{c|}{} 
& OSrCIR \cite{tang2025reason} $_{(CVPR'2025)}$ 
& 30.47 & 31.14 & {\ul 35.03} & {\ul 36.59} & 33.31 \\

\multicolumn{1}{c|}{} 
& \textbf{VMIR-CVI $_{(Ours)}$} 
& \textbf{37.81*} & \textbf{39.16*} & \textbf{41.98*} 
& \textbf{43.07*} & \textbf{40.51*} \\ \hline

\end{tabular}
\end{table*}

\subsubsection{Results on CIRCO}
Following the consistent improvements on CIRR, Table~\ref{tab:compare_circo} further evaluates VMIR-CVI on the more open-domain CIRCO benchmark. VMIR-CVI achieves the best mAP at every cutoff across all three VLP backbones, with statistically significant improvements over the second-best methods. Specifically, its average mAP surpasses OSrCIR~\cite{tang2025reason} by 1.80 and 2.46 points with ViT-B/32 and ViT-L/14, respectively, and outperforms LDRE~\cite{yang2024ldre} by 6.92 points with ViT-G/14. The consistent gains from mAP@5 to mAP@50 demonstrate improvements in both top-ranked precision and the overall ranking of valid targets.
The advantage is particularly pronounced with ViT-G/14. Compared with LDRE~\cite{yang2024ldre} at mAP@5 and mAP@10, and OSrCIR~\cite{tang2025reason} at mAP@25 and mAP@50, VMIR-CVI achieves absolute gains of 6.69, 6.92, 6.95, and 6.48 points, respectively. This indicates that the proposed intent representation can effectively exploit the stronger cross-modal alignment and representation capacity of larger VLP backbones.

Compared with CIRR, which mainly evaluates fine-grained disambiguation among highly similar candidates, CIRCO involves diverse open-domain queries and allows multiple valid targets. The strong CIRCO results therefore demonstrate that VMIR-CVI not only identifies the intended target semantics but also consistently ranks diverse images satisfying the same composed intent. 
 Furthermore, a closer comparison between pseudo-word learning methods and MLLM-based methods further clarifies the source of the improvement. On CIRCO, pseudo-word methods such as SEARLE, Pic2Word, LinCIR, and Context-I2W generally preserve visual evidence from the reference image but may be affected by reference noise and weak target-oriented reformulation. MLLM-based methods such as CIReVL and OSrCIR generate more explicit target descriptions, but they do not explicitly instantiate the retained reference instances in the final query embedding. VMIR-CVI bridges these two directions: VMIR produces a concise target-oriented description, TVID extracts retained visual instances, and HIAF aligns the text-only and instance-augmented representations. Therefore, the results on CIRCO further support the central claim that both VLP-native alignment and correct visual instantiation are needed for robust open-domain zero-shot CIR.

\begin{table*}[t]
\caption{Performance comparison on FashionIQ from fashion products on the validation set. The best and the second-best scores are marked in bold and underlined fonts, respectively. ``*'' denotes p-value$<$0.05 over the second-best baseline.}
\label{tab:fashion}
\vspace{-1em}
\renewcommand\tabcolsep{4pt}
\begin{tabular}{cc|cccccc|cc}
\hline
\multicolumn{2}{c|}{\textbf{FashionIQ →}} & \multicolumn{2}{c}{Shirts} & \multicolumn{2}{c}{Dresses} & \multicolumn{2}{c|}{Top\&Tees} & \multicolumn{2}{c}{\textbf{Average}} \\ \hline
\multicolumn{1}{c|}{Arch} & Method & R@10 & R@50 & R@10 & R@50 & R@10 & R@50 & R@10 & R@50 \\ \hline

\multicolumn{1}{c|}{\multirow{5}{*}{ViT-B/32}} & SEARLE \cite{Baldrati_2023_ICCV} $_{(ICCV'2023)}$ & 24.44 & 41.61 & 18.54 & 39.51 & 25.70 & 46.46 & 22.89 & 42.53 \\
\multicolumn{1}{c|}{} & CIReVL \cite{karthik2023vision} $_{(ICLR'2024)}$ & 28.36 & 47.84 & 25.29 & 46.36 & 31.21 & 53.85 & 28.29 & 49.35 \\
\multicolumn{1}{c|}{} & LDRE \cite{yang2024ldre} $_{(SIGIR'2024)}$ & 27.38 & 46.27 & 19.97 & 41.84 & 27.07 & 48.78 & 24.81 & 45.63 \\
\multicolumn{1}{c|}{} & OSrCIR \cite{tang2025reason} $_{(CVPR'2025)}$ & {\ul 31.16} & {\ul 51.13} & {\ul 29.35} & {\ul 50.37} & {\ul 36.51} & {\ul 58.71} & {\ul 32.34} & {\ul 53.40} \\
\multicolumn{1}{c|}{} & \textbf{VMIR-CVI $_{(Ours)}$} & \textbf{33.81*} & \textbf{54.12*} & \textbf{32.37*} & \textbf{53.99*} & \textbf{39.32*} & \textbf{63.74*} & \textbf{35.17*} & \textbf{57.29*} \\ \hline

\multicolumn{1}{c|}{\multirow{12}{*}{ViT-L/14}} & Pic2Word \cite{saito2023pic2word} $_{(CVPR'2023)}$ & 26.20 & 43.60 & 20.00 & 40.20 & 27.90 & 47.40 & 24.70 & 43.70 \\
\multicolumn{1}{c|}{} & SEARLE \cite{Baldrati_2023_ICCV} $_{(ICCV'2023)}$ & 26.89 & 45.58 & 20.48 & 43.13 & 29.32 & 49.97 & 25.56 & 46.23 \\
\multicolumn{1}{c|}{} & LinCIR \cite{gu2024language} $_{(CVPR'2024)}$ & 29.10 & 46.81 & 20.92 & 42.44 & 28.81 & 50.18 & 26.28 & 46.49 \\
\multicolumn{1}{c|}{} & Context-I2W \cite{tang2024context} $_{(AAAI'2024)}$ & 29.70 & 48.60 & 23.10 & 45.30 & 30.60 & 52.90 & 27.80 & 48.90 \\
\multicolumn{1}{c|}{} & CIReVL \cite{karthik2023vision} $_{(ICLR'2024)}$ & 29.49 & 47.40 & 24.79 & 44.76 & 31.36 & 53.65 & 28.55 & 48.57 \\
\multicolumn{1}{c|}{} & LDRE \cite{yang2024ldre} $_{(SIGIR'2024)}$ & 31.04 & 51.22 & 22.93 & 46.76 & 31.57 & 53.64 & 28.51 & 50.54 \\
\multicolumn{1}{c|}{} & MOA \cite{li2025rethinking} $_{(SIGIR'2025)}$ & 31.90 & 50.70 & 25.20 & 48.50 & 33.20 & 54.80 & 30.10 & 51.30 \\
\multicolumn{1}{c|}{} & MLLM-I2W \cite{bao2025mllm} $_{(COLING'2025)}$ & 27.30 & 46.50 & {\ul 29.90} & 48.60 & 33.80 & 55.20 & 30.30 & 50.10 \\
\multicolumn{1}{c|}{} & PrediCIR \cite{tang2025missing} $_{(CVPR'2025)}$ & 31.80 & 52.00 & 25.40 & 49.50 & 33.10 & 55.40 & 30.10 & 52.30 \\
\multicolumn{1}{c|}{} & OSrCIR \cite{tang2025reason} $_{(CVPR'2025)}$ & 33.17 & 52.03 & 29.70 & {\ul 51.81} & {\ul 36.92} & {\ul 59.27} & {\ul 33.26} & {\ul 54.37} \\
\multicolumn{1}{c|}{} & TT-RLDR \cite{zhou2026duplex} $_{(AAAI'2026)}$ & {\ul 33.46} & {\ul 52.65} & 24.86 & 46.62 & 34.52 & 54.77 & 30.95 & 51.35 \\
\multicolumn{1}{c|}{} & \textbf{VMIR-CVI $_{(Ours)}$} & \textbf{36.11*} & \textbf{55.59*} & \textbf{31.33*} & \textbf{54.83*} & \textbf{41.76*} & \textbf{62.93*} & \textbf{36.40*} & \textbf{57.78*} \\ \hline

\multicolumn{1}{c|}{\multirow{4}{*}{ViT-G/14}} & CIReVL \cite{karthik2023vision} $_{(ICLR'2024)}$ & 33.71 & 51.42 & 27.07 & 49.53 & 35.80 & 56.14 & 32.19 & 52.36 \\
\multicolumn{1}{c|}{} & LDRE \cite{yang2024ldre} $_{(SIGIR'2024)}$ & 35.94 & \textbf{58.58} & 26.11 & 51.12 & 35.42 & 56.67 & 32.49 & 55.46 \\
\multicolumn{1}{c|}{} & OSrCIR \cite{tang2025reason} $_{(CVPR'2025)}$ & \textbf{38.65} & 54.71 & {\ul 33.02} & {\ul 54.78} & {\ul 41.04} & {\ul 61.83} & {\ul 37.57} & {\ul 57.11} \\
\multicolumn{1}{c|}{} & \textbf{VMIR-CVI $_{(Ours)}$} & {\ul 38.42} & {\ul 57.46} & \textbf{35.35*} & \textbf{58.70*} & \textbf{43.60*} & \textbf{65.73*} & \textbf{39.12*} & \textbf{60.63*} \\ \hline

\end{tabular}
\end{table*}

\subsubsection{Results on FashionIQ}
Beyond the open-domain evaluations on CIRR and CIRCO, Table~\ref{tab:fashion} evaluates the generalization of VMIR-CVI to FashionIQ, a specialized product-domain benchmark involving fine-grained modifications of garment attributes. With ViT-B/32 and ViT-L/14, VMIR-CVI achieves the best results across all category-specific and average metrics. 

Specifically, it surpasses OSrCIR~\cite{tang2025reason}, the strongest baseline in terms of average performance, by 2.83/3.89 and 3.14/3.41 points in average Recall@10/Recall@50, respectively. These statistically significant gains demonstrate that VMIR-CVI improves both precise target identification and broader retrieval coverage.
The improvements are particularly evident on Tops\&Tees, where retrieval commonly depends on localized details such as sleeve shape, printed patterns, and decorative elements. For example, with ViT-L/14, VMIR-CVI outperforms OSrCIR~\cite{tang2025reason} by 4.84 and 3.66 points at Recall@10 and Recall@50, respectively. This result is consistent with the proposed design: VMIR translates fine-grained modifications into concise, target-oriented descriptions, TVID preserves intent-relevant local visual cues while suppressing irrelevant content, and HIAF integrates the textual intent and visual evidence into a more discriminative query representation.

With ViT-G/14, VMIR-CVI achieves the best performance on both Dresses and Tops\&Tees, as well as the highest average Recall@10 and Recall@50. It improves over OSrCIR~\cite{tang2025reason} by 1.55 and 3.52 points on the two average metrics, while obtaining the second-best results on Shirts. Together with the results on CIRR and CIRCO, these findings show that VMIR-CVI remains effective across both open-domain and specialized product retrieval, validating its ability to model fine-grained, attribute-centric compositions while preserving the correct visual instance.

\subsubsection{\textbf{Answer to RQ1}}
Overall, VMIR-CVI achieves the strongest overall performance on CIRR and CIRCO across the evaluated backbones, and obtains the best average FashionIQ results under ViT-B/32, ViT-L/14, and ViT-G/14. The consistent gains on CIRR and CIRCO demonstrate its effectiveness in open-domain CIR, where the intended target must be inferred by jointly modeling semantic modifications and reference-specific visual evidence. The results on FashionIQ further confirm its generalization to a specialized product domain involving fine-grained attribute changes, despite slightly lower performance on the Shirts category under ViT-G/14. Collectively, these results validate that VLP-aligned target-oriented reasoning, intent-consistent visual instantiation, and hybrid-modal fusion are effective and jointly enable accurate and robust zero-shot CIR.

\subsection{Ablation Studies (RQ2)} 
To investigate the effectiveness of the key components in VMIR-CVI (RQ2), we conduct systematic ablation studies using ViT-L/14 as the backbone. Following a coarse-to-fine protocol, we first remove each major component from the complete pipeline to quantify its overall contribution. We then perform component-level analyses by replacing each module with basic alternatives. Specifically, VMIR is compared with variants without multi-step CoT reasoning and with a standard multi-step CoT strategy. TVID is replaced by global visual features or a basic attention mechanism. And HIAF is evaluated by removing CIPE, the target-oriented textual representation, or the Combiner. This design enables a systematic assessment of both the necessity of each component and the effectiveness of its specific formulation.

\subsubsection{Overall contribution of VMIR, TVID, and HIAF}

\begin{table}[t]
\caption{Ablation studies on the public FashionIQ dataset under ViT-L/14 VLP backbone.}\label{tab:ablation_fiq}
\begin{tabular}{c|ccccccccc}
\hline
                           & \multicolumn{9}{c}{FashionIQ}                                                                                            \\ \cline{2-10} 
                           & \multicolumn{3}{c|}{Shirts}                & \multicolumn{3}{c|}{Dress}                 & \multicolumn{3}{c}{Tops\&Tees} \\ \cline{2-10} 
\multirow{-3}{*}{Variants} & R@10  & R@50  & \multicolumn{1}{c|}{Avg.}  & R@10  & R@50  & \multicolumn{1}{c|}{Avg.}  & R@10     & R@50     & Avg.     \\ \hline
\rowcolor[HTML]{EFEFEF} 
\textbf{VMIR-CVI} &
  \textbf{36.11} &
  \textbf{55.59} &
  \multicolumn{1}{c|}{\cellcolor[HTML]{EFEFEF}\textbf{45.85}} &
  \textbf{31.33} &
  \textbf{54.83} &
  \multicolumn{1}{c|}{\cellcolor[HTML]{EFEFEF}\textbf{43.08}} &
  \textbf{41.76} &
  \textbf{62.93} &
  \textbf{52.35} \\ \hline
-TVID                      & 35.87 & 54.17 & \multicolumn{1}{c|}{45.02} & 28.56 & 52.85 & \multicolumn{1}{c|}{40.71} & 39.11    & 60.33    & 49.72    \\
-TVID-VMIR                 & 34.25 & 53.58 & \multicolumn{1}{c|}{43.92} & 24.99 & 49.38 & \multicolumn{1}{c|}{37.19} & 36.56    & 58.34    & 47.45    \\
-TVID-VMIR-HIAF            & 27.43 & 45.04 & \multicolumn{1}{c|}{36.24} & 19.19 & 39.91 & \multicolumn{1}{c|}{29.55} & 29.27    & 50.13    & 39.70    \\ \hline
\end{tabular}
\end{table}

\begin{table}[t]
\caption{Ablation studies on the public CIRCO and CIRR dataset under ViT-L/14 VLP backbone.}\label{tab:ablation_circo_cirr}
\begin{tabular}{c|ccccc|ccccc}
\hline
                           & \multicolumn{5}{c|}{CIRCO}            & \multicolumn{5}{c}{CIRR}              \\ \cline{2-11} 
                           & \multicolumn{5}{c|}{mAP@k}            & \multicolumn{5}{c}{Recall@k}          \\ \cline{2-11} 
\multirow{-3}{*}{Variants} & k=5   & k=10  & k=25  & k=50  & Avg.  & k=1   & k=5   & k=10  & k=50  & Avg.  \\ \hline
\rowcolor[HTML]{EFEFEF} 
\textbf{VMIR-CVI} &
  \textbf{26.48} &
  \textbf{27.72} &
  \textbf{30.26} &
  \textbf{31.36} &
  \textbf{28.96} &
  \textbf{30.00} &
  \textbf{60.22} &
  \textbf{70.82} &
  \textbf{89.71} &
  \textbf{62.69} \\ \hline
-TVID                      & 24.19 & 25.61 & 28.07 & 29.09 & 26.74 & 28.60 & 57.52 & 68.84 & 88.48 & 60.86 \\
-TVID-VMIR                 & 21.06 & 22.10 & 24.21 & 25.17 & 23.14 & 23.74 & 51.23 & 64.07 & 85.95 & 56.25 \\
-TVID-VMIR-HIAF            & 8.68  & 9.18  & 10.28 & 10.90 & 9.76  & 22.12 & 48.70 & 60.27 & 85.08 & 54.04 \\ \hline
\end{tabular}
\end{table}

Table~\ref{tab:ablation_fiq} and Table~\ref{tab:ablation_circo_cirr} present the overall ablation results under the ViT-L/14 backbone. We observe a consistent performance degradation as the proposed components are progressively removed, indicating that VMIR, TVID, and HIAF all make non-trivial contributions to VMIR-CVI.

Removing TVID leads to clear drops across FashionIQ, CIRCO, and CIRR, with the largest degradation appearing on fine-grained fashion categories and the open-domain CIRCO benchmark. This verifies the importance of disentangling intent-consistent visual instance cues from the reference image, especially when the retrieval target is determined by local attributes or when the scene contains multiple potentially distracting objects.

When VMIR is further removed, the performance decreases more substantially. This suggests that VMIR is not merely a prompt variant, but provides essential retrieval-oriented semantic guidance. By generating concise target descriptions together with explicit retained-instance cues, VMIR helps bridge the gap between the composed query and the native VLP retrieval space, thereby supporting the subsequent visual instantiation and fusion stages.

The largest degradation is observed after removing HIAF, particularly on CIRCO. This demonstrates that simply obtaining textual intent and visual instance cues is insufficient; they must be coherently aligned and fused into a unified query embedding. HIAF plays this role by integrating the VLP-aligned textual representation from VMIR with the disentangled visual evidence from TVID, enabling the final query to better match target images in the frozen VLP space.

Overall, the stepwise decline from the full VMIR-CVI model to its ablated variants provides strong evidence for the complementarity of the three modules. VMIR contributes target-oriented intent reasoning, TVID provides fine-grained and noise-suppressed visual instantiation, and HIAF transforms these heterogeneous signals into a coherent VLP-compatible retrieval representation. These results validate the effectiveness of the proposed design in constructing visually grounded and retrieval-aligned composed queries.
 \begin{figure}[t] 
    	\centering
    	\includegraphics[width=1\linewidth]{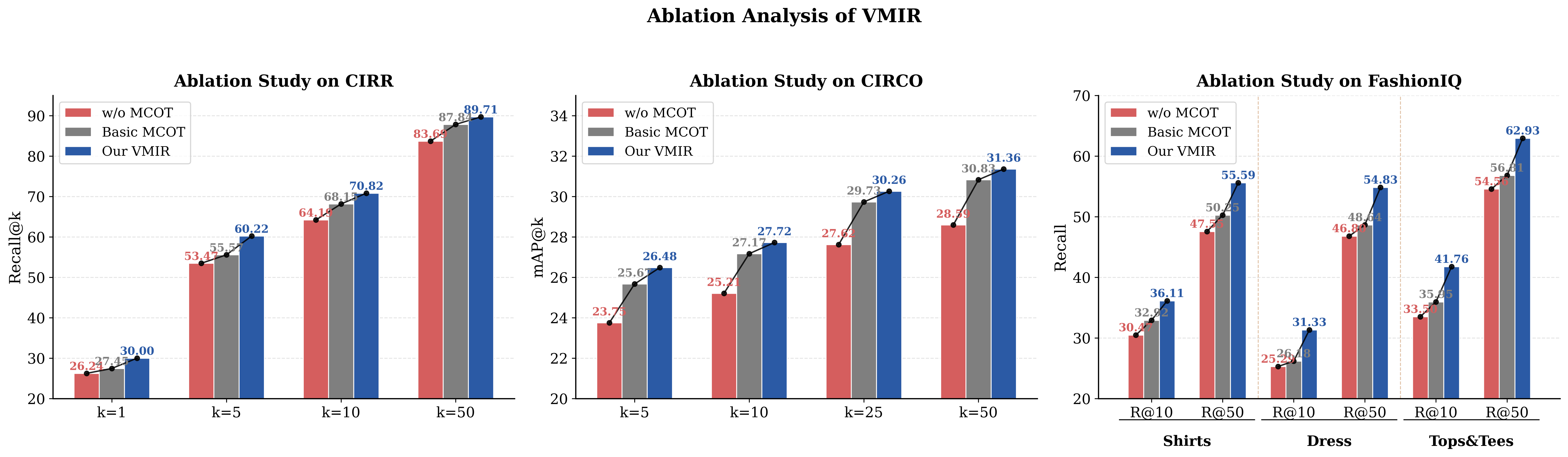}
    	\vspace{-2em}
     \caption{Ablation study of the proposed VMIR on CIRR, CIRCO and FashionIQ.
     }
    	\label{abl-vmir-cot}    
    \end{figure}

\subsubsection{Ablation on VMIR}
Figure~\ref{abl-vmir-cot} evaluates the effectiveness of VMIR by comparing three reasoning strategies: w/o MCoT, Basic MCoT, and the proposed VMIR. The w/o MCoT variant uses a simple direct reasoning prompt, while Basic MCoT introduces multi-step decomposition over the reference image, modification text, and target intent. In contrast, VMIR further incorporates VLP-aligned query exemplars and retained-instance reasoning, encouraging the MLLM to generate concise, target-oriented, and retrieval-compatible descriptions.

Across all three benchmarks, Basic MCoT consistently improves over w/o MCoT, showing that decomposing the composed query is beneficial for identifying what should be modified and what should be preserved. However, VMIR achieves further improvements over Basic MCoT on almost all metrics, with particularly clear gains on FashionIQ and CIRR. For example, compared with Basic MCoT, VMIR improves the average FashionIQ scores by around 4--6 points across the three categories, and increases the average CIRR score by nearly 3 points. These results indicate that generic multi-step reasoning alone is insufficient for zero-shot CIR; the generated query must also be well aligned with the expression style and retrieval space of the frozen VLP backbone.
The advantage of VMIR is especially evident on FashionIQ, where retrieval often depends on compact attribute-object compositions such as color, sleeve length, pattern, or garment shape, etc. By reformulating the composed intent into a concise target-centric description, VMIR helps the VLP text encoder focus on the correct modified attributes while preserving the relevant reference instance. 

Overall, these results verify that VMIR is not merely a prompt variant, but an effective VLP-aligned intent reasoning module. It provides both retrieval-compatible target descriptions and explicit retained-instance cues, thereby establishing a stronger semantic foundation for subsequent visual instantiation and hybrid-modal fusion.

\begin{table}[t]
\caption{Retrieval performance (\%) of VMIR with different MLLM-based reasoning models on FashionIQ.}
\label{tab:MLLM_fiq}
\begin{tabular}{c|ccccccccc}
\hline
\multirow{3}{*}{MLLMS} & \multicolumn{9}{c}{FashionIQ}                                                                                            \\ \cline{2-10} 
                       & \multicolumn{3}{c|}{Shirts}                & \multicolumn{3}{c|}{Dress}                 & \multicolumn{3}{c}{Tops\&Tees} \\ \cline{2-10} 
                       & R@10  & R@50  & \multicolumn{1}{c|}{Avg.}  & R@10  & R@50  & \multicolumn{1}{c|}{Avg.}  & R@10     & R@50     & Avg.     \\ \hline
Qwen3-VL-2B\footnotemark[1]            & 31.55 & 48.97 & \multicolumn{1}{c|}{40.26} & 26.82 & 49.03 & \multicolumn{1}{c|}{37.93} & 35.80    & 55.94    & 45.87    \\
Qwen3-VL-8B\footnotemark[2]            & 32.29 & 50.54 & \multicolumn{1}{c|}{41.42} & 28.11 & 50.37 & \multicolumn{1}{c|}{39.24} & 37.23    & 56.60    & 46.92    \\
Qwen3-VL-30B-A3B\footnotemark[3] &
  \textbf{36.11} &
  \textbf{55.59} &
  \multicolumn{1}{c|}{\textbf{45.85}} &
  \textbf{31.33} &
  \textbf{54.83} &
  \multicolumn{1}{c|}{\textbf{43.08}} &
  \textbf{41.76} &
  \textbf{62.93} &
  \textbf{52.35} \\ \hline
\end{tabular}
\end{table}

\begin{table}[t]
\caption{Retrieval performance (\%) of VMIR with different MLLM-based reasoning models on CIRCO and CIRR.}
\label{tab:MLLM_circo_cirr}
\begin{tabular}{c|ccccc|ccccc}
\hline
\multirow{3}{*}{MLLMS} & \multicolumn{5}{c|}{CIRCO}            & \multicolumn{5}{c}{CIRR}              \\ \cline{2-11} 
                       & \multicolumn{5}{c|}{mAP@k}            & \multicolumn{5}{c}{Recall@k}          \\ \cline{2-11} 
                       & k=5   & k=10  & k=25  & k=50  & Avg.  & k=1   & k=5   & k=10  & k=50  & Avg.  \\ \hline
Qwen3-VL-2B\footnotemark[1]             & 24.24 & 25.43 & 28.05 & 29.05 & 26.69 & 25.18 & 53.74 & 66.53 & 87.18 & 58.16 \\
Qwen3-VL-8B\footnotemark[2]             & 25.85 & 27.09 & 29.59 & 30.62 & 28.29 & 28.55 & 58.89 & 69.66 & 89.35 & 61.61 \\
Qwen3-VL-30B-A3B\footnotemark[3]  &
  \textbf{26.48} &
  \textbf{27.72} &
  \textbf{30.26} &
  \textbf{31.36} &
  \textbf{28.96} &
  \textbf{30.00} &
  \textbf{60.22} &
  \textbf{70.82} &
  \textbf{89.71} &
  \textbf{62.69} \\ \hline
\end{tabular}
\end{table}

\footnotetext[1]{Qwen3-VL-2B-Instruct: https://huggingface.co/Qwen/Qwen3-VL-2B-Instruct}
\footnotetext[2]{Qwen3-VL-8B-Instruct: https://huggingface.co/Qwen/Qwen3-VL-8B-Instruct}
\footnotetext[3]{Qwen3-VL-30B-A3B-Instruct: https://huggingface.co/Qwen/Qwen3-VL-30B-A3B-Instruct}

\subsubsection{Sensitivity to different MLLMs}

Table~\ref{tab:MLLM_fiq} and Table~\ref{tab:MLLM_circo_cirr} report the performance of VMIR when different Qwen3-VL reasoning models are used. The results show a clear scaling trend. Qwen3-VL-2B obtains 40.26/37.93/45.87/26.69/58.16, Qwen3-VL-8B improves to 41.42/39.24/46.92/28.29/61.61, and Qwen3-VL-30B-A3B reaches 45.85/43.08/52.35/28.96/62.69.

From 2B to 8B, the absolute improvements are 1.16, 1.31, 1.05, 1.60, and 3.45 points. From 8B to 30B-A3B, the additional improvements are 4.43, 3.84, 5.43, 0.67, and 1.08 points. This pattern suggests that stronger MLLMs provide better intent parsing and retained-instance selection, especially on FashionIQ categories where fine-grained attribute language is important. At the same time, the 2B and 8B variants remain competitive, indicating that VMIR-CVI does not rely exclusively on the largest MLLM. The framework can still benefit from smaller reasoners because TVID and HIAF provide additional visual grounding and retrieval-space alignment.

The relatively smaller 8B-to-30B gain on CIRCO also provides an important insight. Once target descriptions and retained instances reach a reasonable quality level, further improvement on open-domain retrieval may be constrained by visual grounding and the final VLP embedding space rather than language reasoning alone. This supports the design choice of combining VMIR with TVID and HIAF instead of relying only on an MLLM-generated text query.

\begin{table}[t]
\caption{Ablation study of the proposed TVID on FashionIQ.}\label{tab:tvid_ablation_fiq}
\begin{tabular}{c|ccccccccc}
\hline
                        & \multicolumn{9}{c}{FashionIQ}                                                                                            \\ \cline{2-10} 
                        & \multicolumn{3}{c|}{Shirts}                & \multicolumn{3}{c|}{Dress}                 & \multicolumn{3}{c}{Tops\&Tees} \\ \cline{2-10} 
\multirow{-3}{*}{MLLMS} & R@10  & R@50  & \multicolumn{1}{c|}{Avg.}  & R@10  & R@50  & \multicolumn{1}{c|}{Avg.}  & R@10     & R@50     & Avg.     \\ \hline
Global Attention        & 35.87 & 54.17 & \multicolumn{1}{c|}{45.02} & 28.56 & 52.85 & \multicolumn{1}{c|}{40.71} & 39.11    & 60.33    & 49.72    \\
Basic Attention         & 34.69 & 54.66 & \multicolumn{1}{c|}{44.68} & 28.71 & 52.11 & \multicolumn{1}{c|}{40.41} & 38.70    & 59.71    & 49.21    \\
\rowcolor[HTML]{EFEFEF} 
Our TVID &
  \textbf{36.11} &
  \textbf{55.59} &
  \multicolumn{1}{c|}{\cellcolor[HTML]{EFEFEF}\textbf{45.85}} &
  \textbf{31.33} &
  \textbf{54.83} &
  \multicolumn{1}{c|}{\cellcolor[HTML]{EFEFEF}\textbf{43.08}} &
  \textbf{41.76} &
  \textbf{62.93} &
  \textbf{52.35} \\ \hline
\end{tabular}
\end{table}

\begin{table}[t]
\caption{Ablation study of the proposed TVID on CIRCO and CIRR.}\label{tab:tvid_ablation_circo_cirr}
\begin{tabular}{c|ccccc|ccccc}
\hline
                        & \multicolumn{5}{c|}{CIRCO}            & \multicolumn{5}{c}{CIRR}              \\ \cline{2-11} 
                        & \multicolumn{5}{c|}{mAP@k}            & \multicolumn{5}{c}{Recall@k}          \\ \cline{2-11} 
\multirow{-3}{*}{MLLMS} & k=5   & k=10  & k=25  & k=50  & Avg.  & k=1   & k=5   & k=10  & k=50  & Avg.  \\ \hline
Global Attention        & 24.19 & 25.61 & 28.07 & 29.09 & 26.74 & 28.60 & 57.52 & 68.84 & 88.48 & 60.86 \\
Basic Attention         & 25.60 & 26.64 & 29.20 & 30.19 & 27.91 & 28.53 & 57.06 & 68.77 & 88.75 & 60.78 \\
\rowcolor[HTML]{EFEFEF} 
Our TVID &
  \textbf{26.48} &
  \textbf{27.72} &
  \textbf{30.26} &
  \textbf{31.36} &
  \textbf{28.96} &
  \textbf{30.00} &
  \textbf{60.22} &
  \textbf{70.82} &
  \textbf{89.71} &
  \textbf{62.69} \\ \hline
\end{tabular}
\end{table}

\subsubsection{Ablation on TVID}
After validating the importance of VLP-aligned reasoning in VMIR, we further examine whether correct visual instantiation is necessary for composed retrieval. Table~\ref{tab:tvid_ablation_fiq} and Table~\ref{tab:tvid_ablation_circo_cirr} compare the proposed TVID with two simpler reference-image modeling strategies, i.e. Global Attention and Basic Attention approaches. The former relies on coarse reference-image cues, while the latter introduces a generic attention mechanism. Neither explicitly disentangles intent-relevant visual instances from irrelevant objects, background regions, or co-occurring visual content.

TVID consistently outperforms both Global Attention and Basic Attention across FashionIQ, CIRCO, and CIRR. On FashionIQ, TVID raises the category-wise average scores to 45.85/43.08/52.35, improving over the stronger baseline by up to 2.63 points. The gains are more evident on Dresses and Tops\&Tees, suggesting that coarse visual aggregation is insufficient for fine-grained attribute composition. On CIRCO and CIRR, TVID also improves the average scores to 28.96 and 62.69, surpassing the stronger alternative by 1.05 and 1.83 points, respectively.
These results reveal the limitations of the two simpler strategies. Global Attention retains broad reference-image context, but its holistic aggregation tends to introduce background clutter, unchanged content, and other intent-irrelevant cues. Basic Attention provides partial localization and performs better than Global Attention on CIRCO, but it remains inferior to TVID across all settings, indicating that generic attention may still focus on visually salient yet modification-irrelevant regions. Therefore, reference-image information should not be injected into the composed query in a coarse or weakly localized manner. Instead, effective zero-shot CIR requires identifying and preserving the visual instances that are truly consistent with the modification intent. The proposed TVID addresses this requirement by constructing cleaner and more discriminative instance-level visual evidence, thereby providing more reliable visual grounding for the final hybrid-modal query representation.

Overall, these results verify the effectiveness of TVID as a training-free visual instance disentanglement module. Rather than treating the reference image as undifferentiated visual tokens, TVID preserves intent-consistent instance-level evidence while suppressing irrelevant visual noise, thereby providing more reliable visual grounding for the subsequent HIAF module.

\begin{table}[t]
\caption{Ablation study of the proposed HIAF on FashionIQ.}\label{tab:hiaf_ablation_fiq}
\begin{tabular}{c|ccccccccc}
\hline
                        & \multicolumn{9}{c}{FashionIQ}                                                                                            \\ \cline{2-10} 
                        & \multicolumn{3}{c|}{Shirts}                & \multicolumn{3}{c|}{Dress}                 & \multicolumn{3}{c}{Tops\&Tees} \\ \cline{2-10} 
\multirow{-3}{*}{MLLMS} & R@10  & R@50  & \multicolumn{1}{c|}{Avg.}  & R@10  & R@50  & \multicolumn{1}{c|}{Avg.}  & R@10     & R@50     & Avg.     \\ \hline
w/o CIPE                & 34.10 & 53.09 & \multicolumn{1}{c|}{43.60} & 29.15 & 51.91 & \multicolumn{1}{c|}{40.53} & 38.45    & 59.05    & 48.75    \\
w/o Target Texts        & 30.86 & 49.95 & \multicolumn{1}{c|}{40.41} & 26.23 & 48.64 & \multicolumn{1}{c|}{37.44} & 33.71    & 56.55    & 45.13    \\
w/o Combiner            & 33.86 & 54.02 & \multicolumn{1}{c|}{43.94} & 28.81 & 51.71 & \multicolumn{1}{c|}{40.26} & 37.63    & 59.51    & 48.57    \\
\rowcolor[HTML]{EFEFEF} 
our HIAF &
  \textbf{36.11} &
  \textbf{55.59} &
  \multicolumn{1}{c|}{\cellcolor[HTML]{EFEFEF}\textbf{45.85}} &
  \textbf{31.33} &
  \textbf{54.83} &
  \multicolumn{1}{c|}{\cellcolor[HTML]{EFEFEF}\textbf{43.08}} &
  \textbf{41.76} &
  \textbf{62.93} &
  \textbf{52.35} \\ \hline
\end{tabular}
\end{table}

\begin{table}[t]
\caption{Ablation study of the proposed HIAF on CIRCO and CIRR.}\label{tab:hiaf_ablation_circo_cirr}
\begin{tabular}{c|ccccc|ccccc}
\hline
                        & \multicolumn{5}{c|}{CIRCO}            & \multicolumn{5}{c}{CIRR}              \\ \cline{2-11} 
                        & \multicolumn{5}{c|}{mAP@k}            & \multicolumn{5}{c}{Recall@k}          \\ \cline{2-11} 
\multirow{-3}{*}{MLLMS} & k=5   & k=10  & k=25  & k=50  & Avg.  & k=1   & k=5   & k=10  & k=50  & Avg.  \\ \hline
w/o CIPE                & 25.59 & 26.65 & 29.20 & 30.20 & 27.91 & 26.63 & 56.39 & 68.60 & 88.63 & 60.06 \\
w/o Target Texts        & 18.09 & 18.94 & 20.89 & 21.75 & 19.92 & 22.75 & 48.15 & 59.95 & 82.02 & 53.22 \\
w/o Combiner            & 24.87 & 26.20 & 28.67 & 29.80 & 27.39 & 29.04 & 58.84 & 71.06 & 90.05 & 62.25 \\
\rowcolor[HTML]{EFEFEF} 
our HIAF &
  \textbf{26.48} &
  \textbf{27.72} &
  \textbf{30.26} &
  \textbf{31.36} &
  \textbf{28.96} &
  \textbf{30.00} &
  \textbf{60.22} &
  \textbf{70.82} &
  \textbf{89.71} &
  \textbf{62.69} \\ \hline
\end{tabular}
\end{table}

\subsubsection{Ablation on HIAF}
After validating VMIR and TVID, we further analyze how HIAF integrates the reasoned textual intent and the disentangled visual instance cues. Table~\ref{tab:hiaf_ablation_fiq} and Table~\ref{tab:hiaf_ablation_circo_cirr} report the ablation results of three key designs in HIAF: Context-aware Instance Pseudo-word Embedding strategy (CIPE), target-oriented textual reasoned descriptions, and the adaptive Combiner.

Removing CIPE consistently degrades performance across FashionIQ, CIRCO, and CIRR. In particular, the average scores drop by 2--4 points on the FashionIQ categories and by 2.63 points on CIRR. This indicates that visual instance cues should not be inserted into the query as isolated or context-free pseudo tokens. Instead, their contextual placement within the target-oriented description is important for preserving the sentence structure while grounding the corresponding visual concepts. Such context-aware grounding makes the instance-augmented query more compatible with the native text space of the frozen VLP backbone.

The variant without target texts suffers the largest degradation, especially on CIRCO and CIRR, where the average scores decrease by 9.04 and 9.47 points, respectively. This confirms that the MLLM-reasoned target description provides the semantic backbone of the composed query. Visual instance cues alone cannot fully express compositional intent, since the modification text may involve attribute changes, relation shifts, object removal, or newly introduced context that is not directly present in the reference image. Therefore, correct visual instantiation should complement, rather than replace, VLP-aligned target-oriented reasoning.

Removing the Combiner also leads to consistent performance drops, showing that pure textual queries and instance-augmented queries provide complementary retrieval signals. The former preserves global semantic clarity, while the latter enhances visual specificity through retained instance cues. By adaptively fusing these two representations, the Combiner prevents the final query from being dominated by either overly generic textual semantics or overly local visual details.

Overall, these results demonstrate the effectiveness of HIAF in constructing a coherent hybrid-modal intent representation. CIPE provides better context-aware visual instantiation cues, reasoned target texts ensure retrieval-aligned semantic intent, and the Combiner adaptively balances textual and visual evidence. Together, they enable VMIR-CVI to better align composed queries with the frozen VLP retrieval space, validating the necessity of hybrid-modal intent alignment and fusion for robust zero-shot CIR.

\noindent \subsubsection{\textbf{Answer to RQ2}}
The ablation studies show that all three components are necessary and complementary. VMIR improves query retrievability by producing VLP-aligned target captions and retained-instance keywords. TVID makes the visual evidence cleaner by disentangling instance-level cues from noisy global reference features. HIAF is the alignment mechanism that converts these heterogeneous signals into a unified embedding compatible with the frozen VLP retrieval space. The largest degradation appears when target text is removed, confirming that zero-shot CIR still relies on explicit semantic intent. However, the consistent improvements from TVID, CIPE, and the Combiner show that visual grounding and hybrid-modal fusion are essential for turning that semantic intent into a more precise retrieval query.



\subsection{Efficiency Analysis (RQ3)}

We further analyze the efficiency of VMIR-CVI from two aspects: pre-training cost and inference cost. For pre-training, VMIR-CVI keeps the VLP backbone frozen and only trains lightweight modules for instance pseudo-word mapping and hybrid-modal fusion, i.e., the Projector $\Psi$ and the Combiner $\mathcal{C}$. Therefore, the additional training overhead is limited. Compared with representative pseudo-word learning methods such as Pic2Word~\cite{saito2023pic2word}, which uses 3M images, and LinCIR~\cite{gu2024language}, which uses 2.47M texts, VMIR-CVI only requires 100K images and around 10GB GPU memory for alignment pre-training. This shows that the proposed method can learn effective hybrid-modal intent alignment with substantially lower pre-training cost, since it does not adapt the VLP backbone but only learns how to integrate disentangled visual instance cues with VLP-aligned textual intent.

For inference, VMIR-CVI does not introduce iterative optimization or additional model fine-tuning. VMIR uses a few-shot VLP-friendly reasoning prompt to generate the target-oriented caption and retained instances in a single MLLM reasoning process, while TVID is training-free and directly operates on frozen VLP features. HIAF then constructs the final hybrid-modal query through lightweight projection and fusion. As a result, the inference latency of VMIR-CVI is around $1.5$s, which is comparable to mainstream MLLM-based zero-shot CIR methods~\cite{tang2025reason}, while achieving stronger retrieval performance. This indicates that the performance gains of VMIR-CVI mainly come from more effective VLP-aligned multimodal intent representation, rather than from heavy backbone tuning or expensive inference-time computation.

\textbf{Answer to RQ3.} Overall, VMIR-CVI achieves strong retrieval performance with minimal alignment pre-training cost and practical inference latency, demonstrating its end-to-end efficiency for zero-shot composed image retrieval.  

\begin{figure*}[t] 
	\centering
	\includegraphics[width=\linewidth]{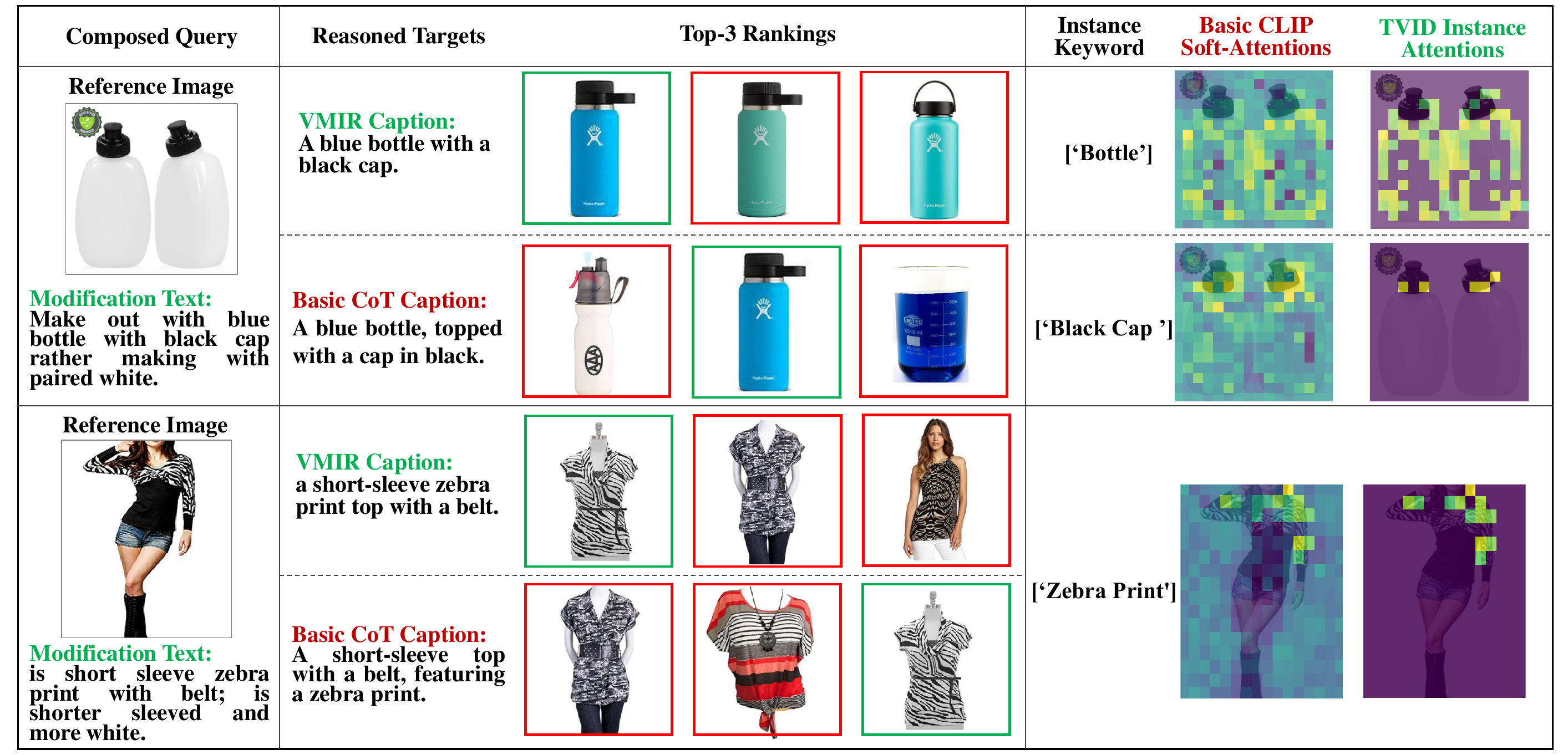}
	\vspace{-1em}
	\caption{Qualitative comparisons between the proposed VMIR and basic CoT reasoning, as well as between TVID and CLIP-based soft-attention mapping. We further present qualitative visualizations of the top-3 retrieval results for different target content reasoning, i.e., with and without VLP-aligned few-shot learning (best viewed in color).} 
	\label{fig:visual_attn_01}
	\vspace{-0.5em}
\end{figure*}  

\subsection{Qualitative Analysis of VMIR and TVID (RQ4)}

To address RQ4, we investigated how VMIR and TVID improve intent reasoning and fine-grained visual disentanglement in the proposed VMIR-CVI by qualitative analysis.
Figure~\ref{fig:visual_attn_01} presents qualitative comparisons between VMIR and basic CoT reasoning, as well as between TVID and CLIP-based soft attention, together with the resulting top-3 ranked retrievals.
From the intent reasoning perspective, VMIR generates concise and VLP-friendly target descriptions, such as ``black cap'' and ``a short-sleeve zebra print'', making it easier for the VLP's encoder to perceive attributed objects more accurately.
From the visual grounding perspective, TVID can disentangle the reference instances more accurately without visual noise, compared with  CLIP-based soft attention.
For instance, given keywords such as ``Bottle'', ``Black Cap'', and ``Zebra Print'', CLIP-based soft-attention exhibits diffuse activations that extend to background regions or co-occurring objects.
In contrast, TVID concentrates responses on the precise visual regions corresponding to the inferred instances, effectively suppressing irrelevant visual cues.
This demonstrates that TVID performs explicit instance-level disentanglement rather than relying on weak global correlations with noise.

\textbf{(Answer to RQ4)} Overall, these qualitative results indicate that VMIR improves the consistency of intent reasoning at the query level with the VLP query paradigm, while TVID enhances fine-grained visual decoupling at the reference level. 
Together, they enable more faithful alignment between composed intent and visual evidence, offering an intuitive explanation for the consistent performance gains observed in zero-shot CIR.

\subsection{Discussion and Limitations}

The above experiments provide several insights into why VMIR-CVI is effective for zero-shot composed image retrieval. First, the strongest and most consistent gains are observed on the open-domain CIRR and CIRCO benchmarks. CIRR requires fine-grained disambiguation among visually similar candidates, while CIRCO evaluates the ranking quality over multiple valid targets in more diverse open-domain scenarios. The strong results on both datasets indicate that VMIR-CVI is able to handle not only precise target identification but also robust intent-level retrieval over diverse visual appearances. This is consistent with our motivation: a composed query should be both semantically explicit and grounded in the correct reference-image instances, rather than relying solely on text-only reasoning or coarse visual pseudo-word injection.

Second, the FashionIQ results further show that VMIR-CVI generalizes to a specialized product domain where retrieval is dominated by fine-grained attribute modifications. VMIR-CVI achieves the best average performance under all evaluated backbones, demonstrating its effectiveness in modeling attribute-centric compositions such as color, shape, pattern, and garment details. Meanwhile, the ViT-G/14 results also reveal category-level variation: VMIR-CVI obtains the best results on Dresses, Tops\&Tees, and the overall average, but is slightly behind the strongest baselines on Shirts. This suggests that although the proposed hybrid-modal intent representation is generally robust, highly constrained product categories may still benefit from category-specific regularities captured by certain baselines.

Third, the ablation studies clarify the complementary roles of the proposed modules. VMIR provides a retrieval-aligned semantic anchor by reformulating the composed query into a concise target-oriented description. TVID improves visual grounding by extracting intent-consistent instance cues while suppressing irrelevant reference-image content. HIAF further integrates these textual and visual signals into a coherent hybrid-modal representation. In particular, the HIAF ablation shows that removing target text causes the largest degradation, confirming that visual instantiation should complement, rather than replace, target-oriented semantics. Similarly, the effectiveness of CIPE indicates that visual pseudo-words are more useful when inserted into semantically appropriate positions instead of being appended as context-free tokens.

Despite these advantages, VMIR-CVI still has several empirical limitations. First, its performance depends on the quality of MLLM-based intent reasoning; inaccurate target descriptions or retained-instance predictions may affect the subsequent visual instantiation and fusion stages. Second, although VMIR-CVI achieves strong average performance on FashionIQ, the category-level results suggest that domain-specific retrieval patterns can still influence the relative advantage of different methods. Third, resolving the exact top-ranked image among near-duplicate candidates remains challenging, especially in cases with subtle visual differences or ambiguous modification texts. These observations motivate future work on more reliable intent reasoning, adaptive instance selection, and more efficient hybrid-modal alignment.

\section{Conclusion and Future Work}
In this paper, we propose VMIR-CVI, a unified framework for zero-shot composed image retrieval that explicitly integrates VLP-aligned multimodal intent reasoning with correct visual instantiation to optimize VLP-aligned multimodal intent representation. To bridge the gap between composed queries and target images under frozen VLP backbones, VMIR-CVI effectively integrates the complementary strengths of MLLM-based intent reasoning and visual pseudo-word representations, while mitigating the inherent limitations of each paradigm.
By guiding MLLMs to generate retrieval-aligned target queries and disentangling fine-grained instance-level visual cues from reference images, VMIR-CVI reconstructs a hybrid-modal intent representation that jointly captures semantic intent and discriminative visual evidence. Furthermore, the proposed hybrid-modal intent alignment and fusion mechanism effectively integrates reasoned textual intent with instantiated visual cues in the VLP embedding space, suppressing intent-irrelevant noise while preserving instance consistency. Extensive experiments on CIRR, CIRCO, and FashionIQ across multiple VLP backbones demonstrate that VMIR-CVI consistently outperforms existing zero-shot CIR methods and establishes new state-of-the-art results.

Despite its effectiveness, VMIR-CVI still has several limitations. First, its performance depends on the quality of MLLM-based intent reasoning, where inaccurate target descriptions or retained-instance predictions may affect subsequent visual instantiation and fusion. Second, the visual instance representations are derived from frozen VLP features, which may be insufficient for highly complex scenes, subtle local attributes, or ambiguous object interactions. In future work, we will explore more adaptive instance alignment, more reliable intent reasoning, and more efficient hybrid-modal fusion mechanisms to further improve robustness and generalization while preserving the efficiency of zero-shot CIR.
\begin{acks}
This work was (partially) supported by the National Natural Science Foundation of China (6260071580, 62506210, 62522210, 62372275, 62472261, 62502282 and 62602383), the Shandong Provincial Natural Science Foundation (ZR2026QC105, ZR2026QC1180), and the Key Scientific and Technological Innovation Program of Shandong Province (No. 2025CXGC010108), the Key R\&D Program of Shandong Province (2025CXGC010112), Qingdao Key Laboratory of Trustworthy Artificial Intelligence,  Shandong Province Key Laboratory of Network Integration Theory and Technology and Shandong Provincial Laboratory for Humanities and Natural Sciences Collaborative Innovation. All content represents the opinion of the authors, which is not necessarily shared or endorsed by their respective employers and/or sponsors.
\end{acks}

\bibliographystyle{ACM-Reference-Format}
\bibliography{sample-base}

\end{document}